\documentclass[epj]{svjour}

\usepackage{graphicx}
\usepackage{xcolor}
\usepackage{amsmath}
\usepackage{amssymb}

\begin{document}
\title{Extreme acoustic anisotropy in crystals visualized by diffraction tensor}
\author{Natalya F. Naumenko\inst{1} \and Konstantin B. Yushkov\inst{1} \and Vladimir Ya. Molchanov\inst{1}}                     
\institute{National University of Science and Technology MISIS, 4 Leninsky prospekt, Moscow 119049, Russia}
\date{Received: date / Revised version: date}
%
\abstract{
Acoustic wave propagation in single crystals, metamaterials and composite structures is a basic mechanism in acoustic, acousto-electronic and acousto-optic devices. Acoustic anisotropy of crystals provides a variety of device performances and application fields, but its role in pre-estimation of achievable device characteristics and location of crystal orientations with desired properties is often underestimated. A geometrical image of acoustic anisotropy can be an important tool in design of devices based on wave propagation in single crystals or combinations of anisotropic materials. We propose a fast and robust method for survey and visualization of acoustic anisotropy based on calculation of the eigenvalues of bulk acoustic wave (BAW) diffraction tensor (curvature of the slowness surface). The stereographic projection of these eigenvalues clearly reveals singular directions of BAW propagation (acoustic axes) in anisotropic media and areas of fast or slow variation of wave velocities. The method is illustrated by application to three crystals of different symmetry used in different types of acoustic devices: paratellurite, lithium niobate, and potassium gadolinium tungstate. The specific features of acoustic anisotropy are discussed for each crystal in terms of their potential application in devices. In addition, we demonstrate that visualization of acoustic anisotropy of lithium niobate helps to find orientations supporting propagation of high-velocity surface acoustic waves.
} 
\maketitle

\section{Introduction}


Recently, there has been a trend of increasing interest in acoustic properties of crystals. This is caused by extensive development of their applications in design of new high-frequency devices. Besides that, acoustic beams in strongly anisotropic crystals are perfect models for fundamental research and demonstration of extreme wave phenomena.
Paradoxically, this theoretically mature field lacks a fast and reliable procedure for comprehensive search of directions suitable for design of specific bulk and surface wave devices. This work was completed to fill this gap. We propose a method for survey and visualization of bulk acoustic wave (BAW) degeneracy and extreme anisotropy directions on stereographic projections of crystals. The method provides {higher robustness and versatility} in finding such directions than commonly used projections of the slowness surface.

Anisotropy of BAW propagation exists in all crystal systems~\cite{Fedorov_eng,Musgrave}. Point symmetry group determines the number of independent elastic and piezoelectric constants and the symmetry elements of the crystal. The quantity of anisotropy among the materials belonging to one symmetry group varies depending on the material constants. For example, acoustic anisotropy among tetragonal crystals varies within the whole order of magnitude~\cite{VoloshinovPolikarpovaDeclercq09}. Large anisotropy of BAW properties enables such phenomena as extremely oblique beam propagation and anomalous backward reflection~\cite{BurovVoloshinovDmitrievPolikarpova11_eng}. A special class of acoustic media with engineered anisotropy are 3D metamaterials~\cite{Deymier,ZigoneanuPopaCummer14,KrodelDaraio16,ZhangWeiCheng17,CuiHensleighYao19}. Acoustic metamaterials and multilayer composite structures designed of anisotropic crystals demonstrate the second level of complexity~\cite{LinHuang11,RollandDupontGazalet14,Naumenko17,NaumenkoNikolay17,Naumenko18,DarinskiiShuvalov18,HagelauerFattingerRuppel18,SahooOttavianoZheng18,SunYu18,WuOudichCao19,Naumenko19}. Deep understanding of BAW anisotropy features of crystalline components is crucial in design of such structures. 

Several following examples illustrate intentional use of strong BAW anisotropy in ultrasonic devices.
Anisotropic propagation of bulk acoustic waves in solids is used in the design of various acousto-optic devices~\cite{GoutzoulisPape,KustersWilsonHammond74,VoloshinovEtal92_eng,KastelikEtal93,AubinSaprielMolchanov04,MolchanovVoloshinovMakarov09_eng,IUS13}. There was an unusual situation. For almost 60 years after the creation of the first laser, more than a thousand articles have been published on the diffraction of light by ultrasonic waves in acoustically anisotropic crystals. Nevertheless, studies of the effect of acoustic anisotropy on the parameters of diffracted optical radiation were performed only fragmentarily and for some special cases of  acousto-optic interaction geometry in selected crystals. A certain exception are some recent works~\cite{BalakshyMantsevich12AJ_eng,NaumenkoMolchanovChizhikovYushkov15,MantsevichBalakshyMolchanovYushkov15,MantsevichMolchanovYushkov17}. {In last years, attention has been focused on crystals of lower symmetry systems, namely orthorhombic and monoclinic ones~\cite{MazurVelikovskiyMazur14,MazurPozhar15_eng,MartynyukTrachKokhanVlokh17,MilkovVoloshinovEtal18_eng,KupreychikBalakshy18,SPIE19_10899}.}
Acousto-optic interaction at GHz-range frequencies in anisotropic microstructures and resonators is also at the edge of applied research~\cite{MachadoCrespoKuznetsov19,ValleBalram19,CaiMahmoudKhan19,TianLiuDong19}.

Acoustic anisotropy plays an important role in investigation of surface acoustic waves (SAW) and optimization of single-crystal substrates for SAW devices: filters, delay lines, sensors, \emph{etc}. The symmetry of a substrate material determines orientations, in which the generalized SAW degenerates into the shear-horizontally (SH) polarized quasi-bulk SAW or sagittally polarized Rayleigh wave solutions, required for application in certain types of SAW filters and sensors. More detailed analysis of acoustic anisotropy enables finding leaky waves suitable for application in high-frequency SAW devices. Though a leaky SAW propagates faster than the slowest BAW and attenuates because of BAW radiation into the substrate, the existence of ``non-leaky'' waves with negligible attenuation and their location on a leaky wave branch~\cite{Naumenko92,Naumenko94}. can be also predicted from analysis of acoustic anisotropy, for example via calculation of ``exceptional wave'' (EW) lines~\cite{AlshitsLothe79a}. These lines comprise one-partial homogeneous solutions of SAW problem and give rise to the branches of quasi-bulk low-attenuated leaky SAWs~\cite{Naumenko95}. In some crystals characterized by strong acoustic anisotropy (quartz, TeO$_2$, Li$_2$B$_4$O$_7$) such leaky SAWs show quasi-longitudinal structure~\cite{AlshitsLyubimovNaumenko85,Naumenko94,Naumenko96}.  They are 1.5--2 times faster than typical SAWs and facilitate manufacturing of high-frequency SAW devices.

Theoretical study of extreme anisotropy directions in crystals is based on search and classification of BAW degeneracy directions. This approach was characterized as ``analyze rather than solve''~\cite{AlshitsLothe04}. It can predict the possible number of acoustic axes in a crystal of given symmetry and additional constraints, e.g. position of acoustic axes relative to symmetry elements of the point group. Degeneracy directions can be found as solutions of a set of algebraic equations. 

A different point of view to the problem is based on computational analysis of BAW properties in a crystal. Following this paradigm, in this work we propose to use curvature of the slowness surface as a quantitative measure of anisotropy. In any point except singular directions, each sheet of the BAW slowness surface is smooth and has two principal curvature values. The choice of this quantity instead of the others has several reasons. First, it is a clear physical interpretation of the slowness surface curvature since its magnitude is proportional to far field divergence of the beam. Second, it does not just allow searching for acoustic axes, but also highlights non singular directions of extreme BAW anisotropy. Those directions can not be found from algebraic approach since they do not satisfy formal degeneracy conditions. Finally, computation of principal curvature is performed by the same procedure for any crystal system, either with or without piezoelectric effect, and requires only the knowledge of elastic moduli and piezoelectric constants.


As a contribution to the field, we analyze the usage of BAW slowness surface curvature tensor to find acoustic axes and directions with strong acoustic anisotropy. We demonstrate that the eigenvalues of this tensor can be used to visualize the position of conical acoustic axes on the stereographic projection of a crystal. Thus, a robust numerical procedure for visualization BAW anisotropy is proposed. This procedure does not impose any limitations on symmetry of the crystal. The proposed method helps to solve some problems related to propagation of acoustic waves in crystals and optimization of crystal orientations for different applications. As the examples, we observe acoustic crystals widely used in acousto-optic and acousto-electronic technology: paratellurite (TeO$_2$), lithium niobate (LNO, LiNbO$_3$), and potassium gadolinium tungstate (KGW, KGd(WO$_4$)$_2$). The examined crystals belong to different crystal systems: tetragonal for paratellurite, trigonal  for LNO, and monoclinic for KGW. Our calculation results fully comply with theoretical predictions on properties of conical BAW axes in crystal and give insight on applications of BAW anisotropy in device physics.

\section{Methodology}
\subsection{Acoustic axes in anisotropic solids}

In this Section we briefly observe the state-of art in theoretical analysis of acoustic axes in crystals.

An acoustic axis is one of specific directions in a crystal, in addition to the longitudinal and transverse normals~\cite{Fedorov_eng}. It is characterized by equal phase velocities of at least two BAWs propagating along the same direction. The theory of acoustic {wave degeneracy} was previously developed by Khatkevich~\cite{Khatkevich77_eng}, Alshits \emph{et al.}~\cite{AlshitsLothe79b,AlshitsShuvalov87_eng,AlshitsLothe06}, Boulanger and Hayes~\cite{BoulangerHayes98}, Shuvalov and Every~\cite{ShuvalovEvery96,ShuvalovEvery97}, Vavry\v{c}uk~\cite{Vavrycuk05}. {The most comprehensive surveys on the topic were made by Alshits and Lothe~\cite{AlshitsLothe04} and Shuvalov~\cite{Shuvalov98}.} Here there are some basic statements about acoustic axes in crystals obtained theoretically, which are important for understanding and correct interpretation of numerical results presented in Section~\ref{sec-res}.

1. The local geometry of the velocity or slowness surface sheets in the vicinity of acoustic axis depends on the type of degeneracy of the acoustic tensor. Tangent degeneracy occurs along a 4-fold symmetry axes in tetragonal or cubic crystals. Conical degeneracy is more typical and exists along a 3-fold symmetry axis, in the planes of crystal symmetry or outside such planes (``off-plane'' acoustic axes). In transversely isotropic media acoustic axes build the cones or the lines on the stereographic projection of propagation directions.

2. The maximum number of acoustic axes in monoclinic, triclinic, and orthorhombic crystals is sixteen. One, five, or nine acoustic axes can exist in a tetragonal crystal. In a trigonal crystal, the total number of acoustic axes is {four, ten, or sixteen}.

3. Acoustic axes were found in all known crystals. However, Alshits and Lothe proved that a model crystal without acoustic axes can exist~\cite{AlshitsLothe04}.

4. The maximum number and types of acoustic axes do not change if piezoelectric effect is present in a crystal but the actual number and locations of acoustic axes generally change. In piezoelectric crystals the ``stiffened'' acoustical tensor, with added piezoelectric contributions, is used instead of pure elastic tensor.

5. Conical acoustic axes are stable to perturbations of elastic or piezoelectric properties. They shift or split rather than disappear when these properties change because of variable growth conditions or after-growth treatment (for example, in chemically reduced LNO wafers with suppressed pyroelectric effect for easier fabrication of SAW devices).

6. Due to degeneracy of acoustical tensor, an infinite number of BAW modes with different polarization vectors can propagate parallel to the acoustic axis. For conical axes, the Poynting vector strikes a cone while propagation direction rotates around the axis. Hence, the phenomena of conical refraction can be observed~\cite{Musgrave,AlshitsLyubimov13_eng,TurpinLoikoKalkandjievMompart16}.

{7. Acoustic axes are associated with singularities in BAW polarization field. Tangent axes have integer topological charge (Poincar\'e index) $N=-1,0,1$. Conical and wedge-point axes have fractional topological charge $N=\pm 1/2$ (or possibly $N=0$ for wedge-point axes). Disappearance and splitting of axes under phase transitions and small perturbations of material constants is constrained by conservation of the topological charge.}

\subsection{BAW diffraction tensor}

The solution of Christoffel equations determines three scalar functions of phase velocities $V$ for any wave normal ${\mathbf n}$. Commonly, the slowness surface $V^{-1}({\mathbf n})$ is used since it is a scaled wave normal surface. It characterizes boundary phenomena (refraction, reflection, and coupling to SAWs)~\cite{Sharma07,VoloshinovPolikarpovaDeclercq09}, band structure in phononic crystals~\cite{Deymier,RollandDupontGazalet14,HernandoLoweCraster15}, and interactions with other waves (e.g. Brillouin scattering of photons)~\cite{Dixon67IEEE,ParyginChirkov75_eng,Laude03}. The normal vector to the slowness surface is the group velocity $\mathbf S$.
The second derivatives of the slowness surface characterize its curvature. It has been shown by Khatkevich~\cite{Khatkevich78_eng} and by Naumenko \emph{et al.}~\cite{NaumenkoPerelomovaBondarenko83_eng} that far-field spreading of a BAW beam is associated with the slowness surface curvature.

\begin{figure}
  \centering
  \includegraphics[width=0.5\columnwidth]{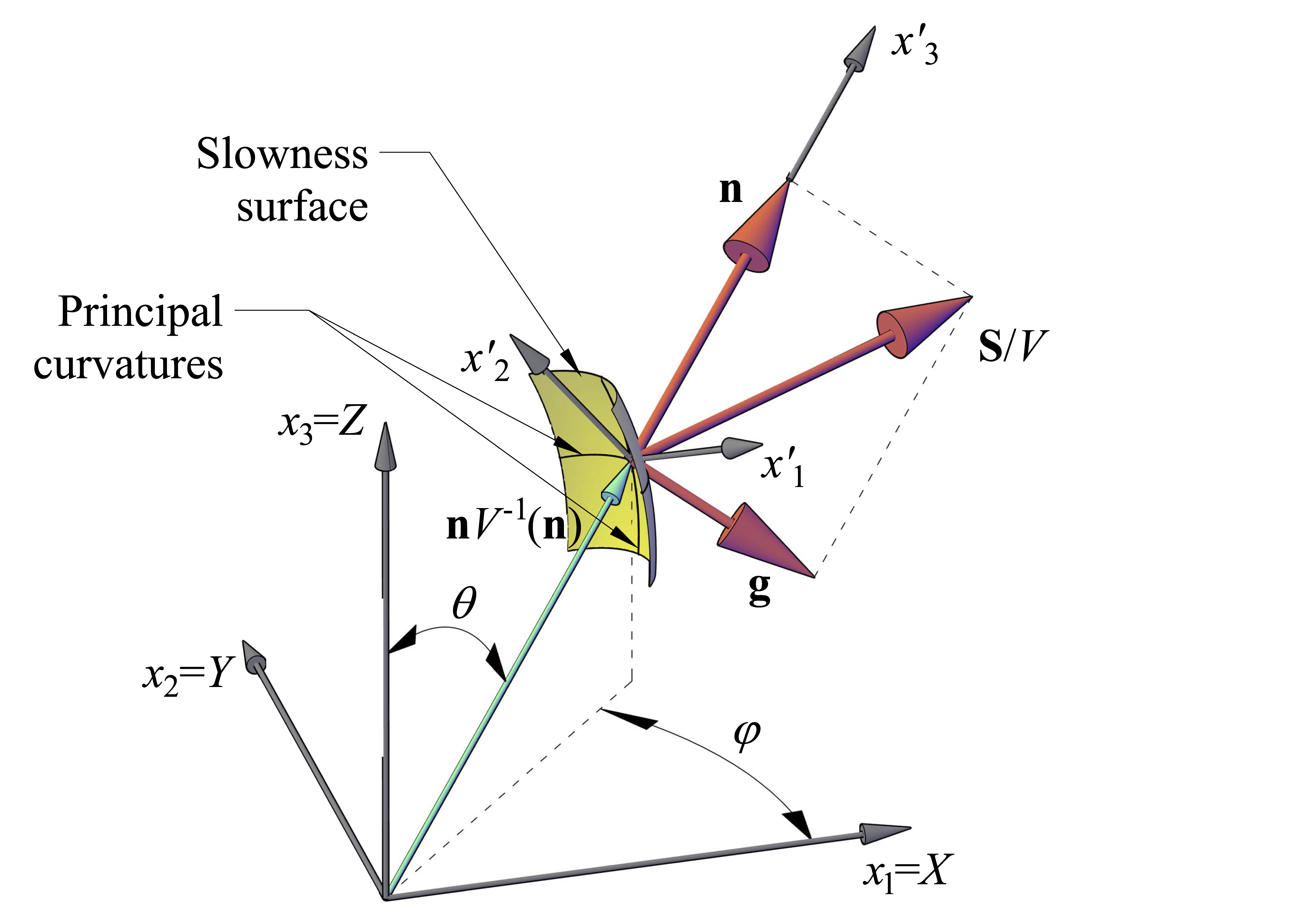}\\
  \caption{Illustration to the definition and of BAW diffraction tensor and its relation to the slowness surface curvature.​}\label{fig-def}
\end{figure}

For calculation of the BAW diffraction tensor the following procedure is used~\cite{NaumenkoPerelomovaBondarenko83_eng,IUS13,NaumenkoMolchanovChizhikovYushkov15}. The normalized transverse component of the group velocity vector $\mathbf S$ is defined as
\begin{equation}\label{eq-Strans}
  \mathbf g = \mathbf S /V - \mathbf n
\end{equation}
A fragment of a slowness surface with the group velocity vector is illustrated in Fig.~\ref{fig-def}. Hereinafter, the wave normal direction is characterized by two Euler angles $\theta$ (between $\mathbf n$ and $x_3$ axis) and $\varphi$ (between projection of $\mathbf n$ on $x_1 x_2$ plane and $x_1$ axis). In a general case, principal curvature planes can be arbitrarily rotated with respect to global coordinate axes $x_1 x_2 x_3$. The local frame of reference $x'_1 x'_2 x'_3$ is associated with the curvature of the slowness surface in the direction of the wave normal $\mathbf n$.

The components of the diffraction tensor are defined as
\begin{equation}\label{eq-Wtens}
  W_{ij} = \delta_{ij} - n_i n_j + g_i g_j + \frac{\partial g_i}{\partial n_j},
\end{equation}
where $\delta_{ij}$ is the Kronecker delta. The diffraction tensor $\widehat W$ is symmetrical and planar, i.e. ${\mathbf n} \widehat W {\mathbf n}=0$. The tensor can be diagonalized to the form
\begin{equation}\label{eq-Wmat}
  \widehat W' =
  \begin{pmatrix}
  w_1 & 0 &0 \\
  0 & w_2 & 0 \\
  0 & 0 & 0 \\
  \end{pmatrix},
\end{equation}
where $w_1, w_2\in\mathbb{R}$ are the eigenvalues. Local axes $x'_1$ and $x'_2$ are selected along the eigenvectors of the tensor $\widehat W$ corresponding to the eigenvalues $w_1$ and $w_2$, so that $w_1>w_2$ for each of the BAW modes. Wave normal $\mathbf n$ is the eigenvector of $\widehat W$ with zero eigenvalue, therefore $x'_3$ axis is {collinear with} $\mathbf n$.

\begin{figure}[!t]
  \centering
  \includegraphics[width=0.5\columnwidth]{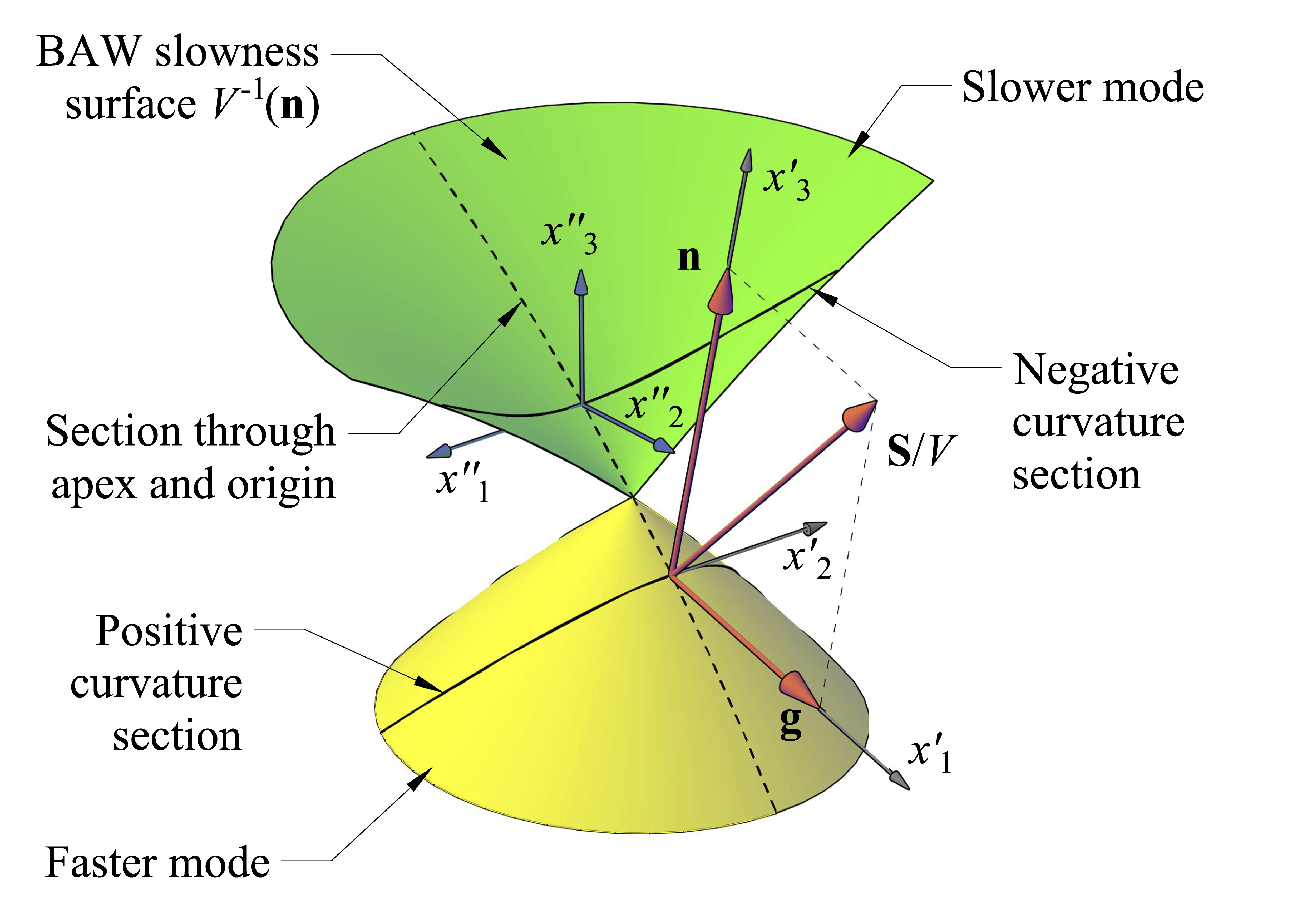}\\
  \caption{General features of the BAW slowness surface at the vicinity of a conical acoustic axis.​}\label{fig-asympt}
\end{figure}

Thus, any direction in a crystal can be characterized with six $w_{\alpha\beta}$ coefficients, where $\alpha=1,2,3$ is the index of the BAW mode and $\beta=1,2$ is the index of the $\widehat W$ tensor eigenvalue. These coefficients are the quantitative estimates of the slowness surface curvature.

The procedure for computation of the BAW diffraction tensor is the following. First, the phase {velocities $V_{\alpha}(\mathbf{n})$ are} calculated from the Christoffel equation. The modes are sorted by their phase velocity, $V_1\geqslant V_2\geqslant V_3$. Second, the components of the group velocity vector {$\mathbf{S}$ are calculated as partial derivatives of $V(\mathbf{n})$ and the transverse component~\eqref{eq-Strans} is found.} Third, the derivatives of $\mathbf g$ are calculated and the diffraction tensor $\widehat W$ is composed according to Eq.~\eqref{eq-Wtens}. Fourth, the eigenvalues of $\widehat W$ are found and sorted.
\begin{figure*}[!t]
  \centering
  \includegraphics[width=0.25\textwidth]{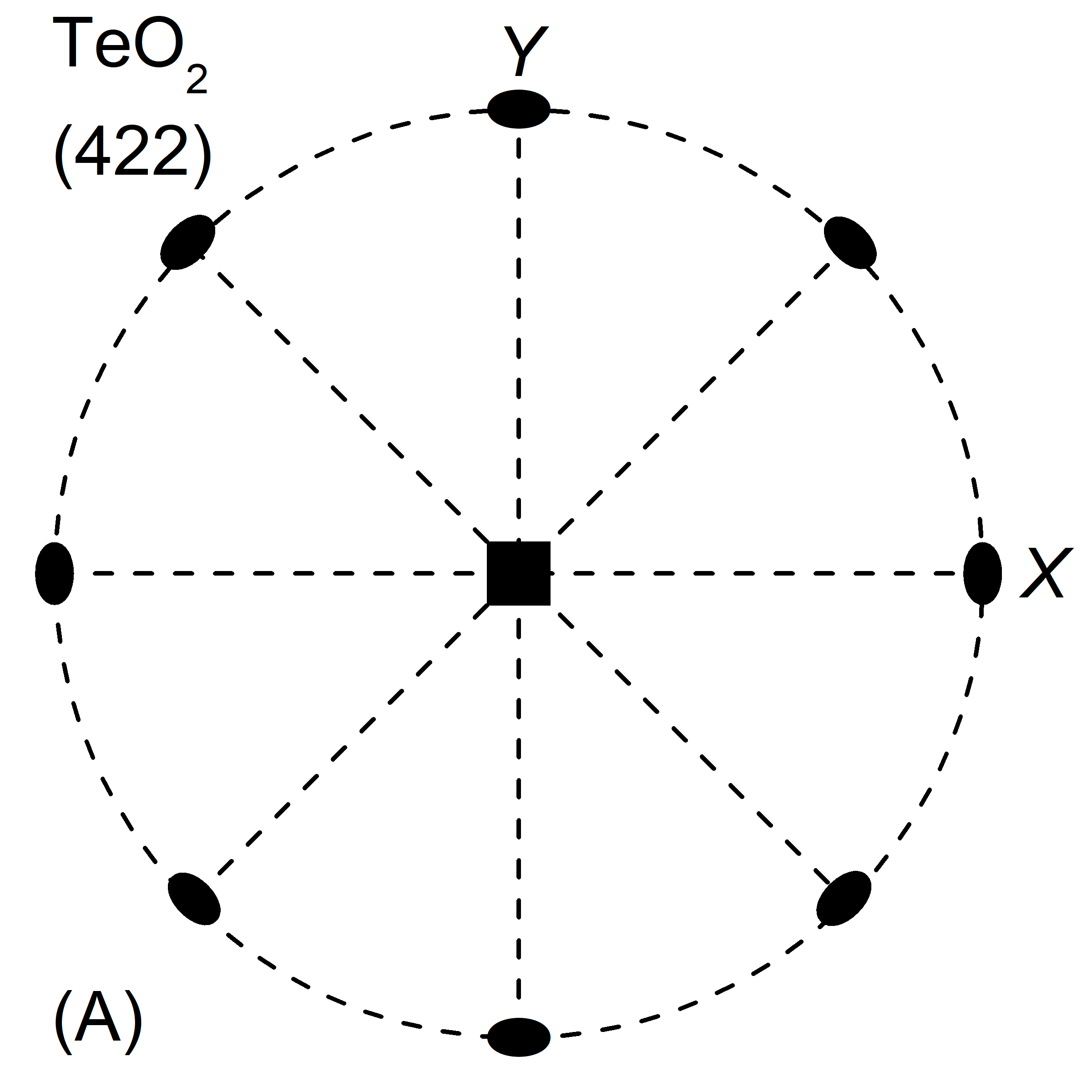}\includegraphics[width=0.25\textwidth]{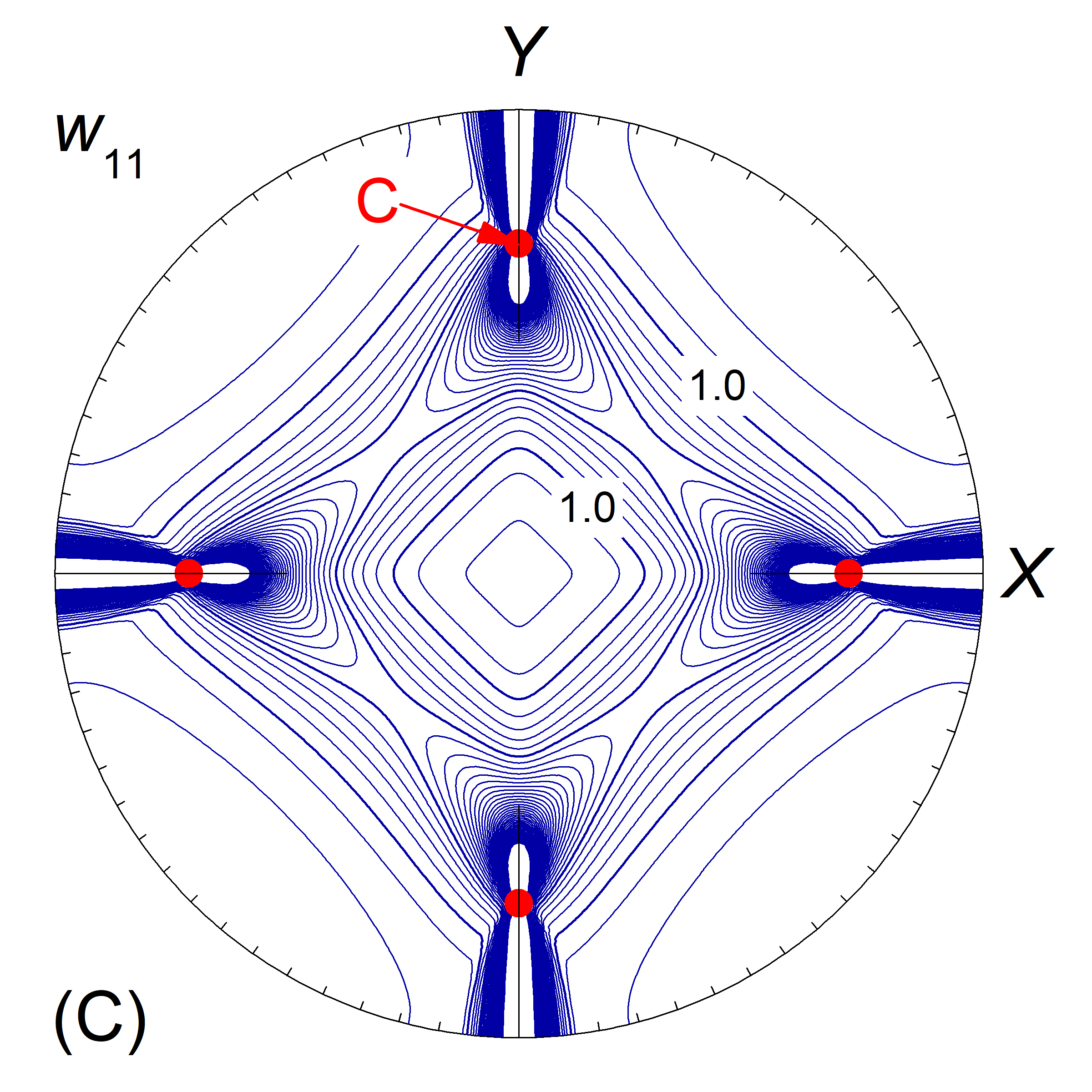}\includegraphics[width=0.25\textwidth]{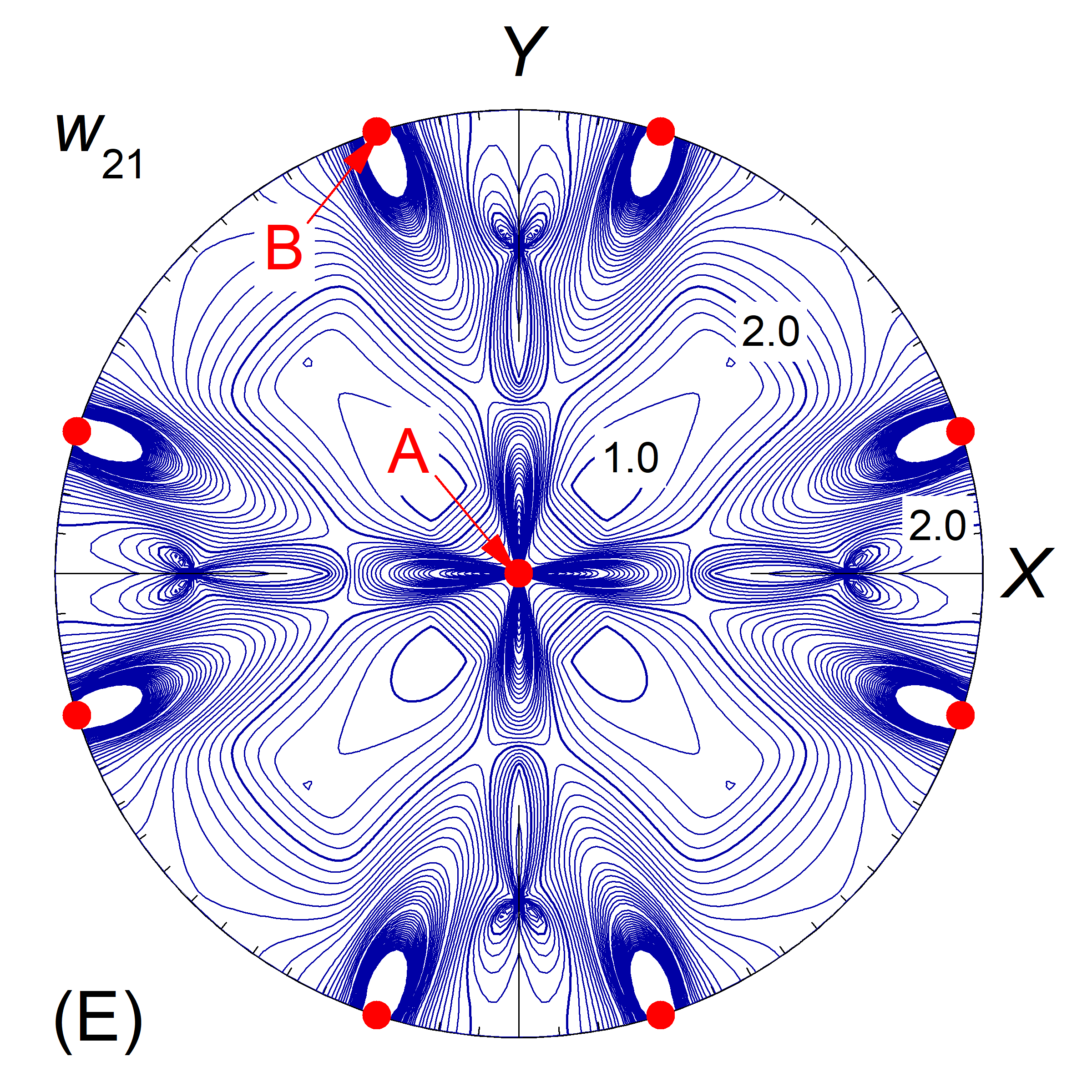}\includegraphics[width=0.25\textwidth]{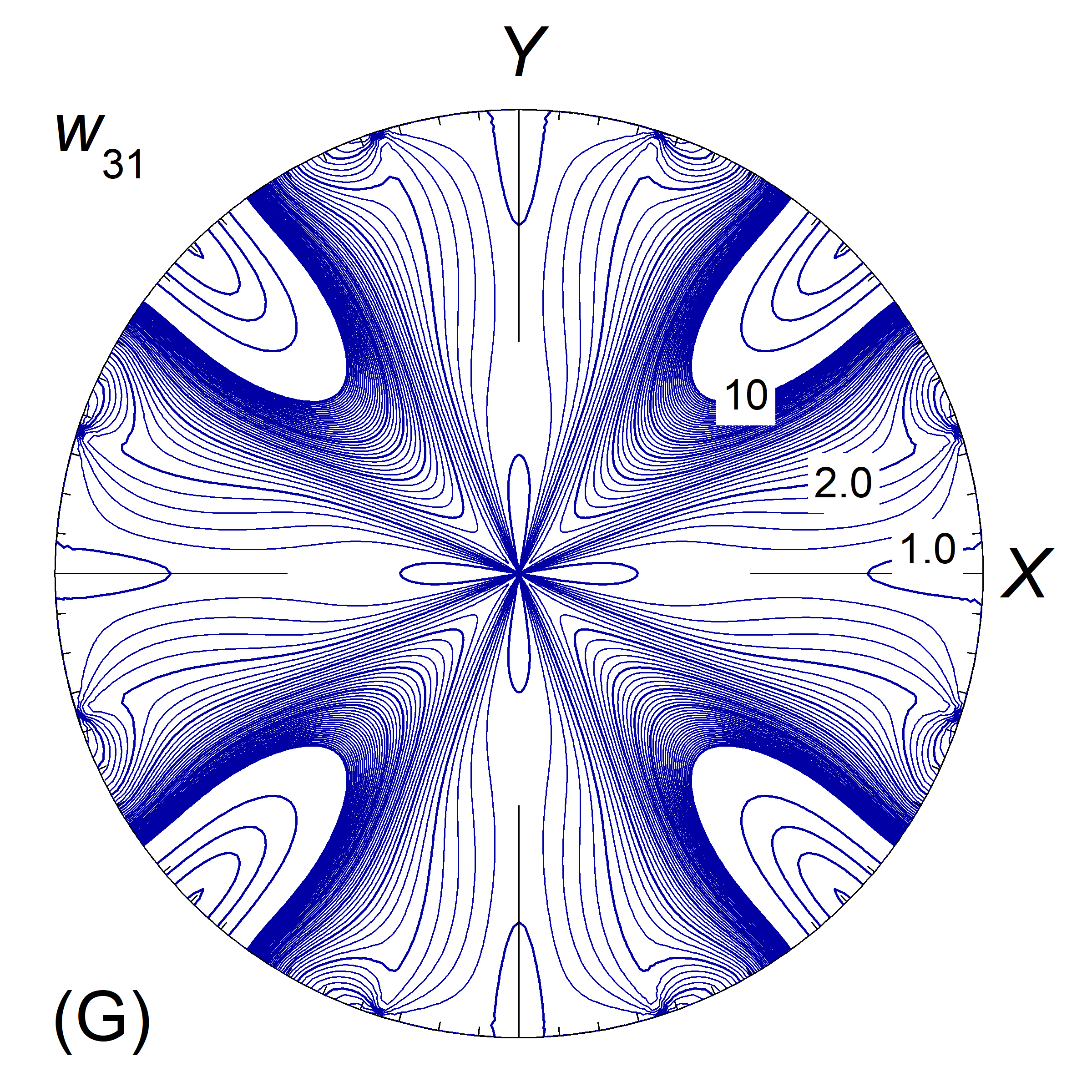}\\
  \includegraphics[width=0.25\textwidth]{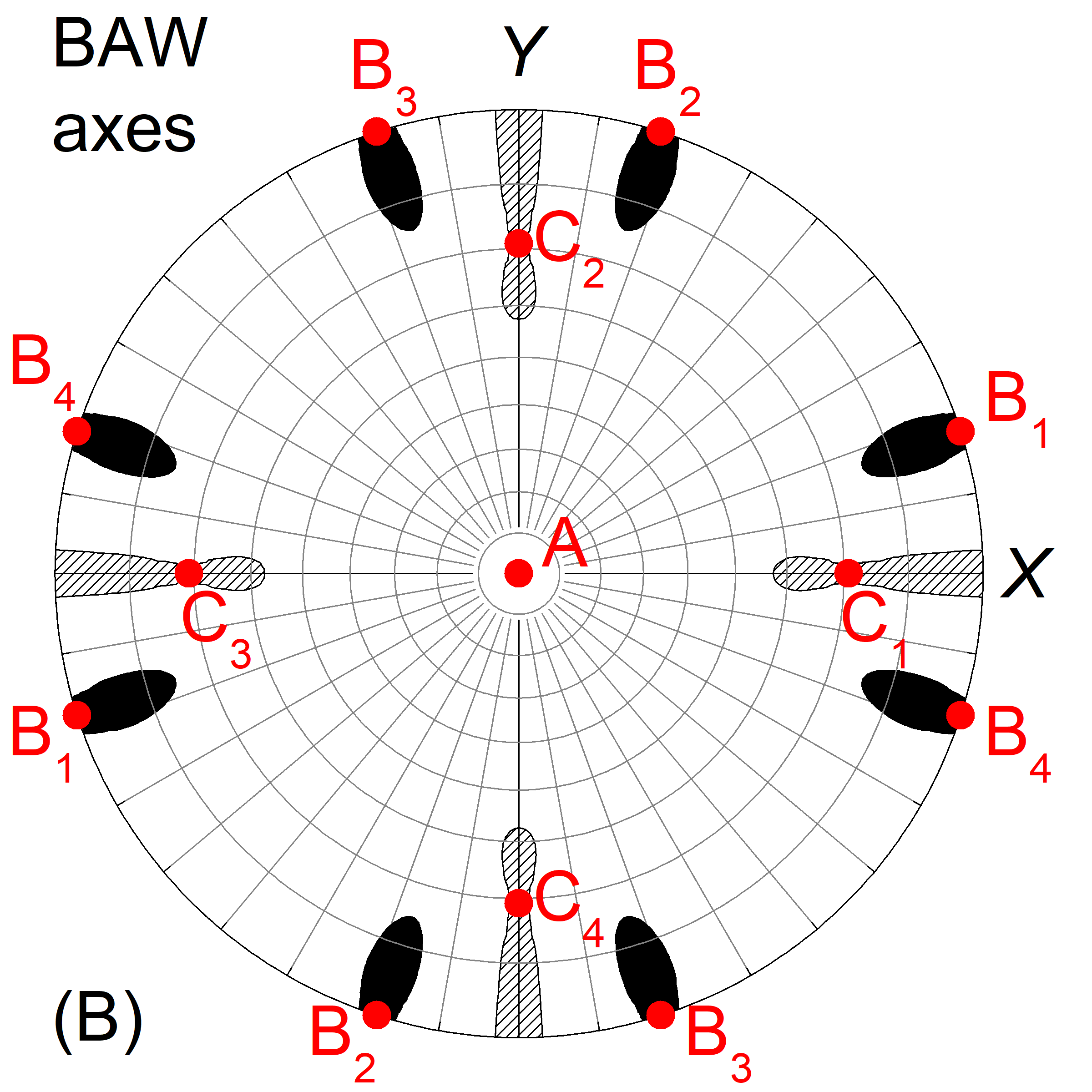}\includegraphics[width=0.25\textwidth]{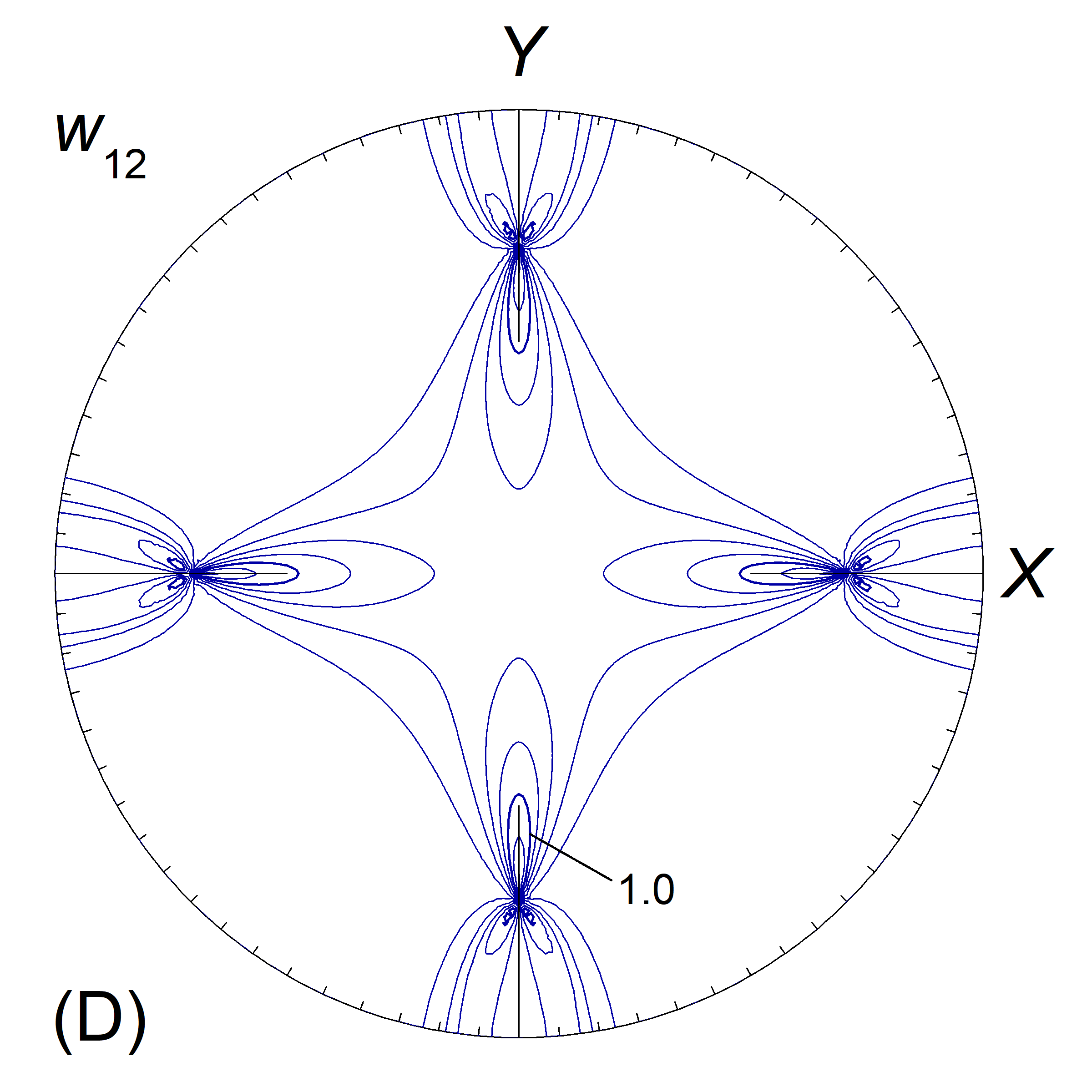}\includegraphics[width=0.25\textwidth]{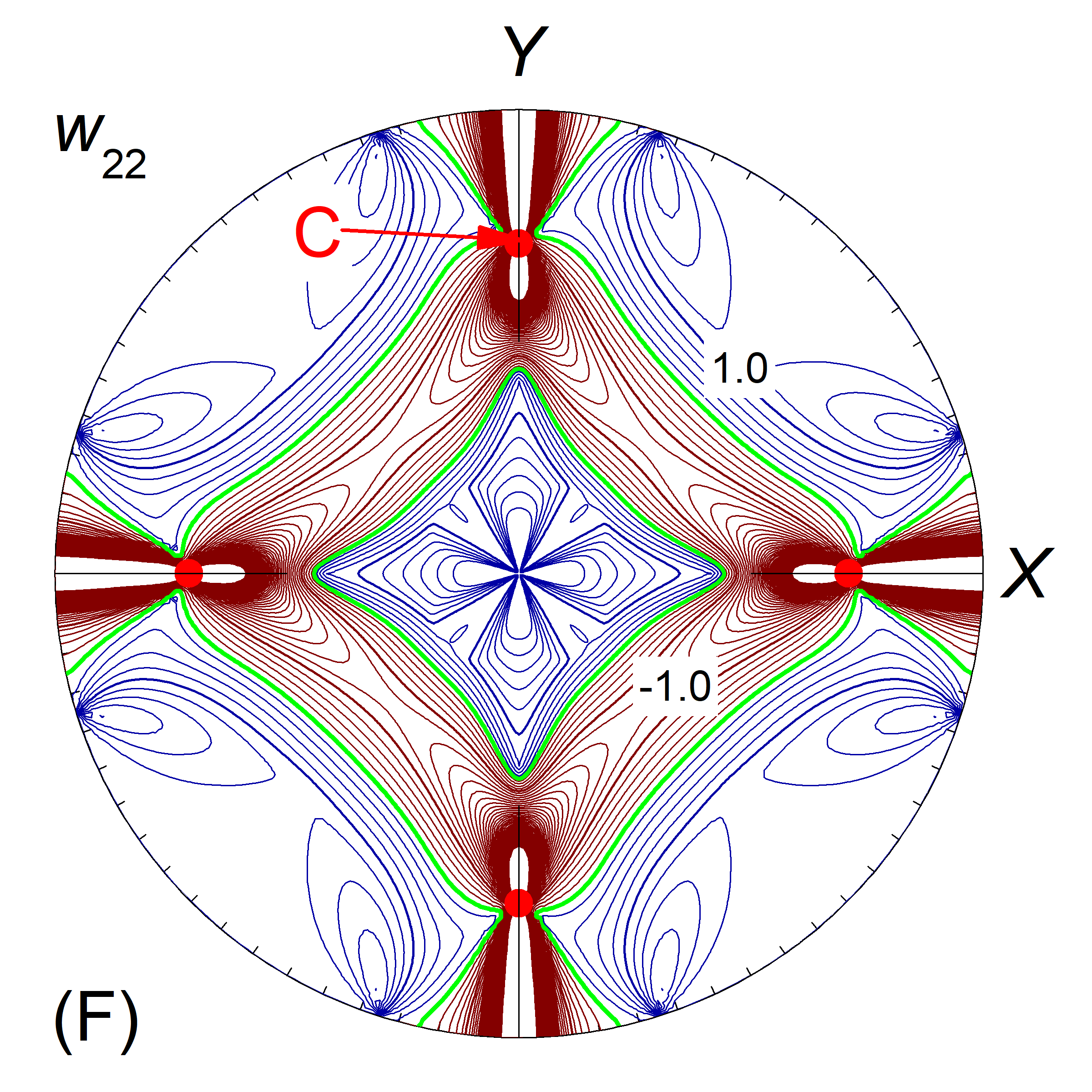}\includegraphics[width=0.25\textwidth]{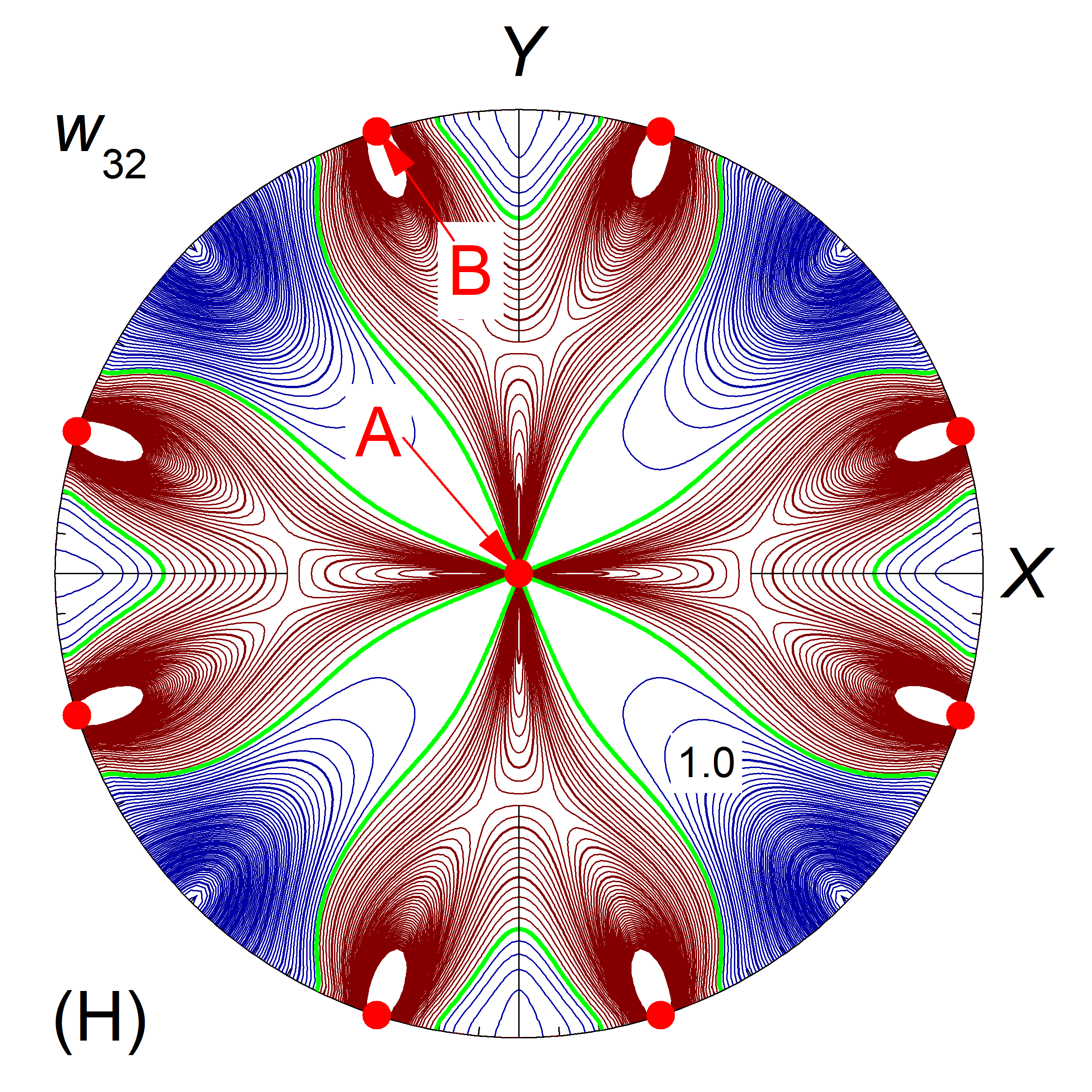}\\
  \caption{Acoustic diffraction tensor in paratellurite crystal: (A) symmetry elements; (B) BAW axes and regions with high anisotropy $w_{11}w_{22}<-40$ (hatch) and $w_{21}w_{32}<-40$ (fill); (C-H) contour plots of BAW diffraction coefficients in stereographic projection (contour spacing is 0.2, for $w_{31}>10$ only major contours with spacing 10 are shown).}\label{fig-TEO2}
\end{figure*}

Six diffraction coefficients are calculated as functions of propagation direction. The calculation procedure takes into account piezoelectric effect as an additional term in the acoustic tensor. To visualize the simulation results we use contour plots of each coefficient $w_{\alpha\beta}$ presented in a stereographic projection on a Wolff net. The numerical results for three crystals of different symmetry are discussed in Section~\ref{sec-res}.

\subsection{Singular points of the diffraction tensor}

A conical axis is an isolated singularity direction on the slowness surface where degeneration of two modes takes place~\cite{AlshitsLothe04}. The geometry of the slowness surface in a conical axis point is shown in Fig.~\ref{fig-asympt}. Curvature of the slowness surface vanishes in the direction of a conical axis. Near the axis, the first eigenvalue for the faster mode is increasing and unbounded while the second eigenvalue for the slower mode is simultaneously decreasing and unbounded:
\begin{equation}\label{eq-infty}
  w_{S2} \rightarrow -\infty,\qquad w_{F1}\rightarrow +\infty,
\end{equation}
{where $F=1,2$ is the index of faster degenerate mode and $S=F+1$ is the index of slower degenerate mode}. Any plane section of the slowness surface through a conical singularity point is a union of smooth intersecting curves. Thus, two other eigenvalues are continuous functions and the equity holds in the degeneracy point~\cite{ShuvalovEvery96}:
\begin{equation}\label{eq}
  w_{S1} = w_{F2}.
\end{equation}

Unbounded growth and decreasing of the diffraction coefficients is visualized as a region with increased density of isolines in contour plots. According to Eq.~\eqref{eq-infty}, unbounded values of the $\widehat W$ tensor eigenvalues simultaneously exist for both of the degenerate modes. Thus, the condition for acoustic degeneration of two faster modes is large positive $w_{11}$ and large in magnitude and negative $w_{22}$. This case is rare in crystal acoustics, but is explicitly illustrated for paratellurite crystals in Sec.~\ref{sec-res-TeO2}. The common case is degeneration of the slower modes. The condition for acoustic degeneration for this case is large positive $w_{21}$ and large in magnitude and negative $w_{32}$.

Besides search for conical axes, contour plots of the diffraction tensor eigenvalues $w_{\alpha\beta}$ are very helpful for visualization regions in crystals with strong acoustic anisotropy. In those directions BAW diffraction coefficients are large in magnitude but finite. Such regions are often observed between conical acoustic axes as demonstrated in Section~\ref{sec-res}. Moreover, such regions may be not neighboring any of the real acoustic axes in the case when degeneracy condition is not fulfilled but the difference of phase velocities of two modes is small. Large diffraction coefficients are also associated with regions of fast BAW velocity changing without mode degeneracy. The case of paratellurite is discussed in detail below.
\begin{figure*}[!t]
  \centering
  \includegraphics[width=0.25\textwidth]{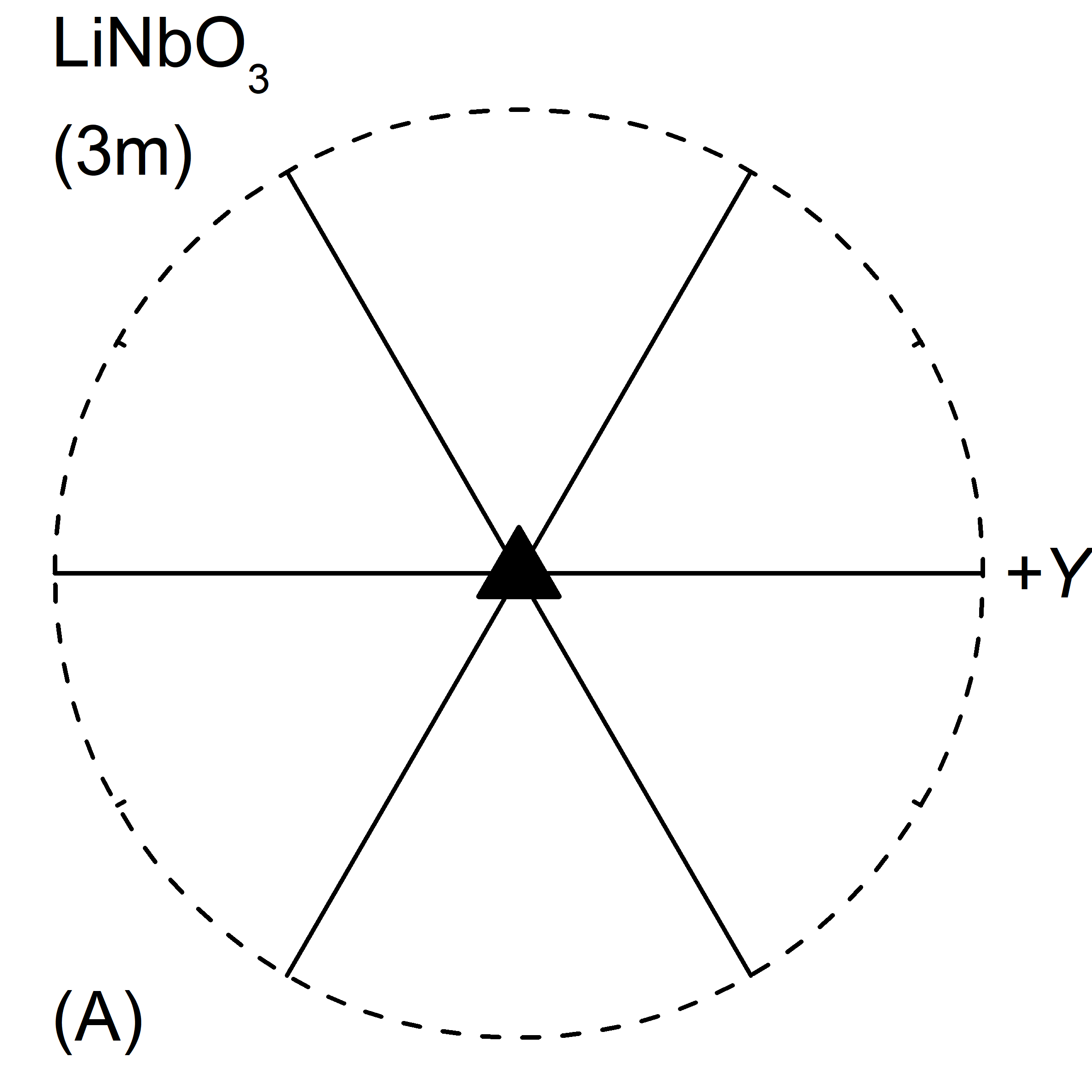}\includegraphics[width=0.25\textwidth]{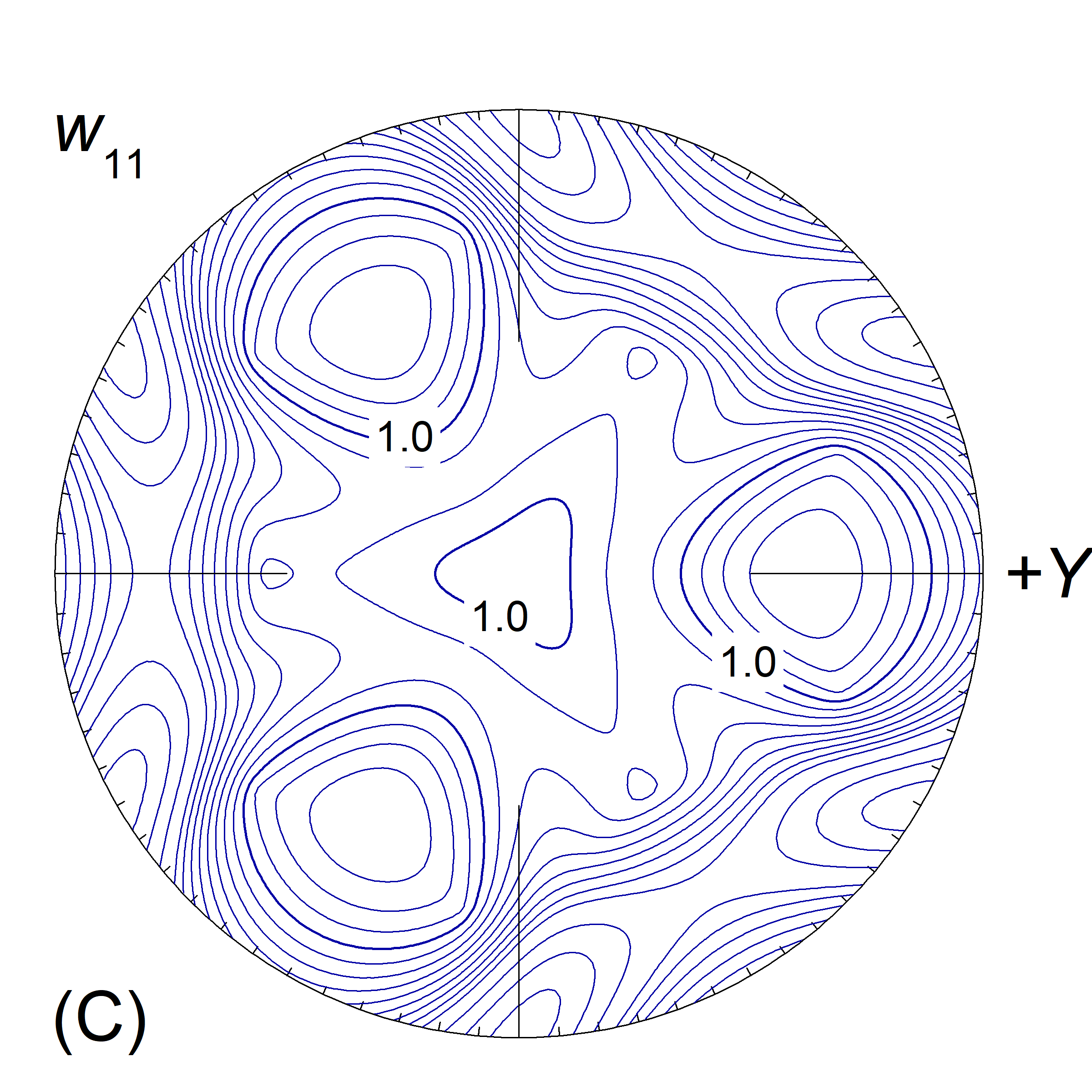}\includegraphics[width=0.25\textwidth]{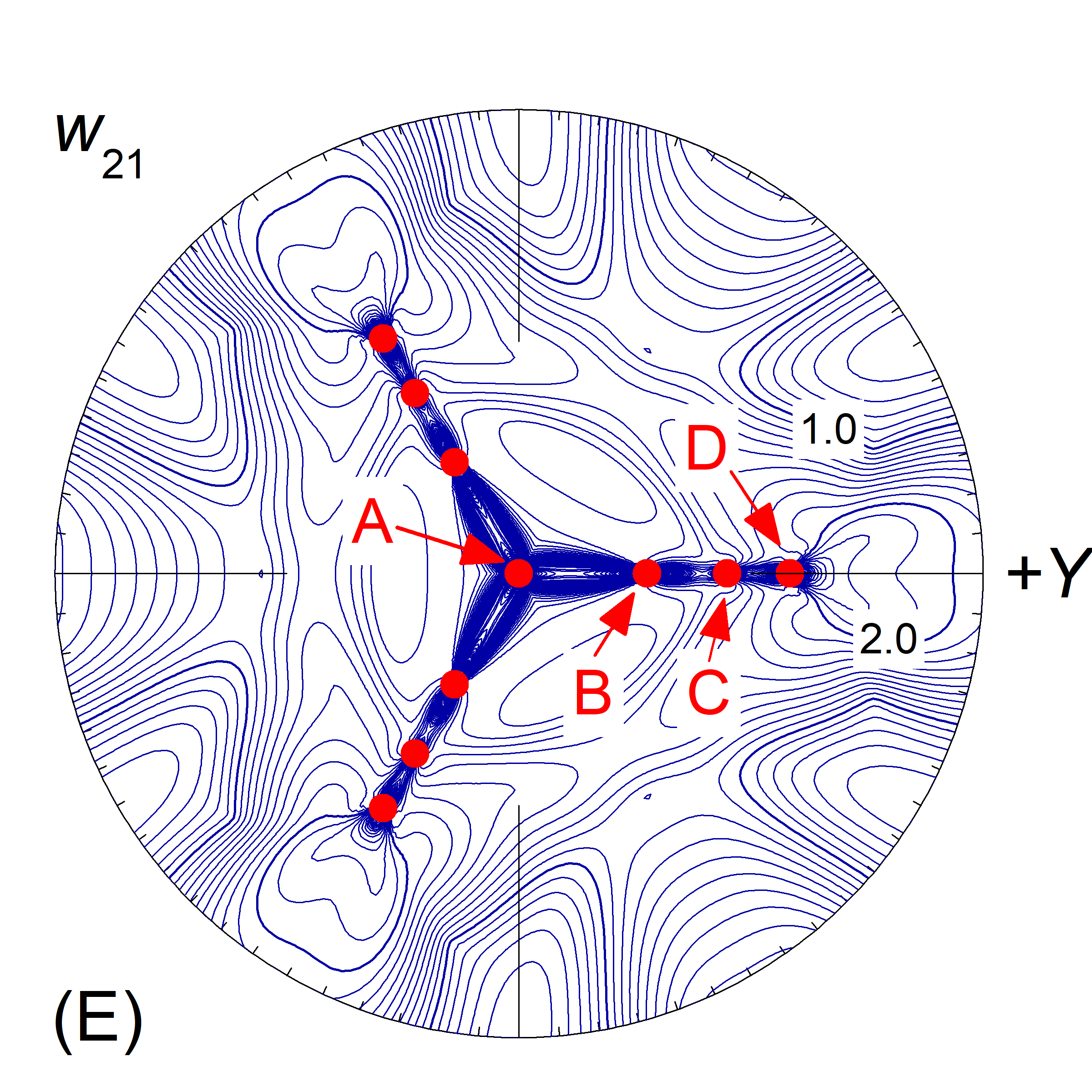}\includegraphics[width=0.25\textwidth]{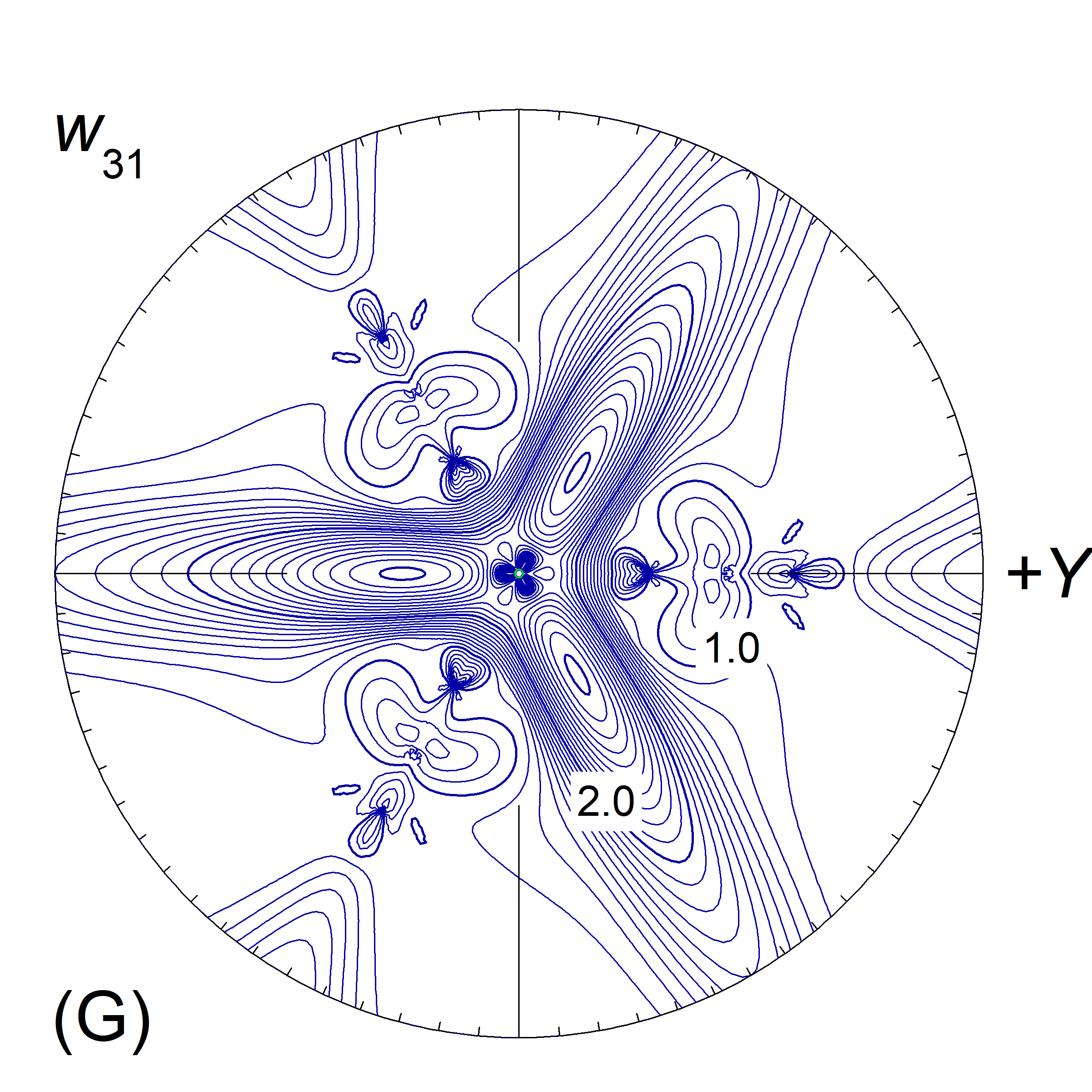}\\
  \includegraphics[width=0.25\textwidth]{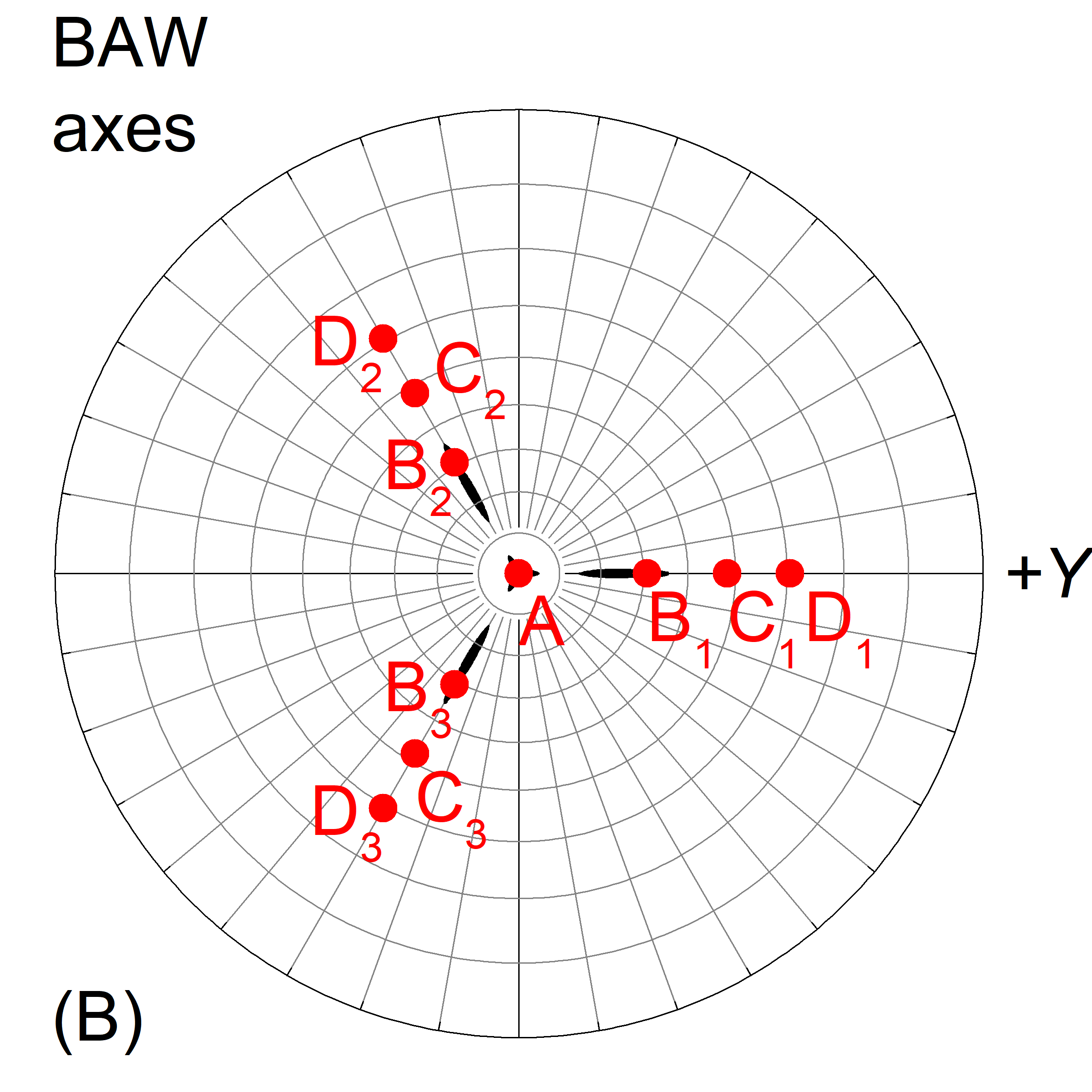}\includegraphics[width=0.25\textwidth]{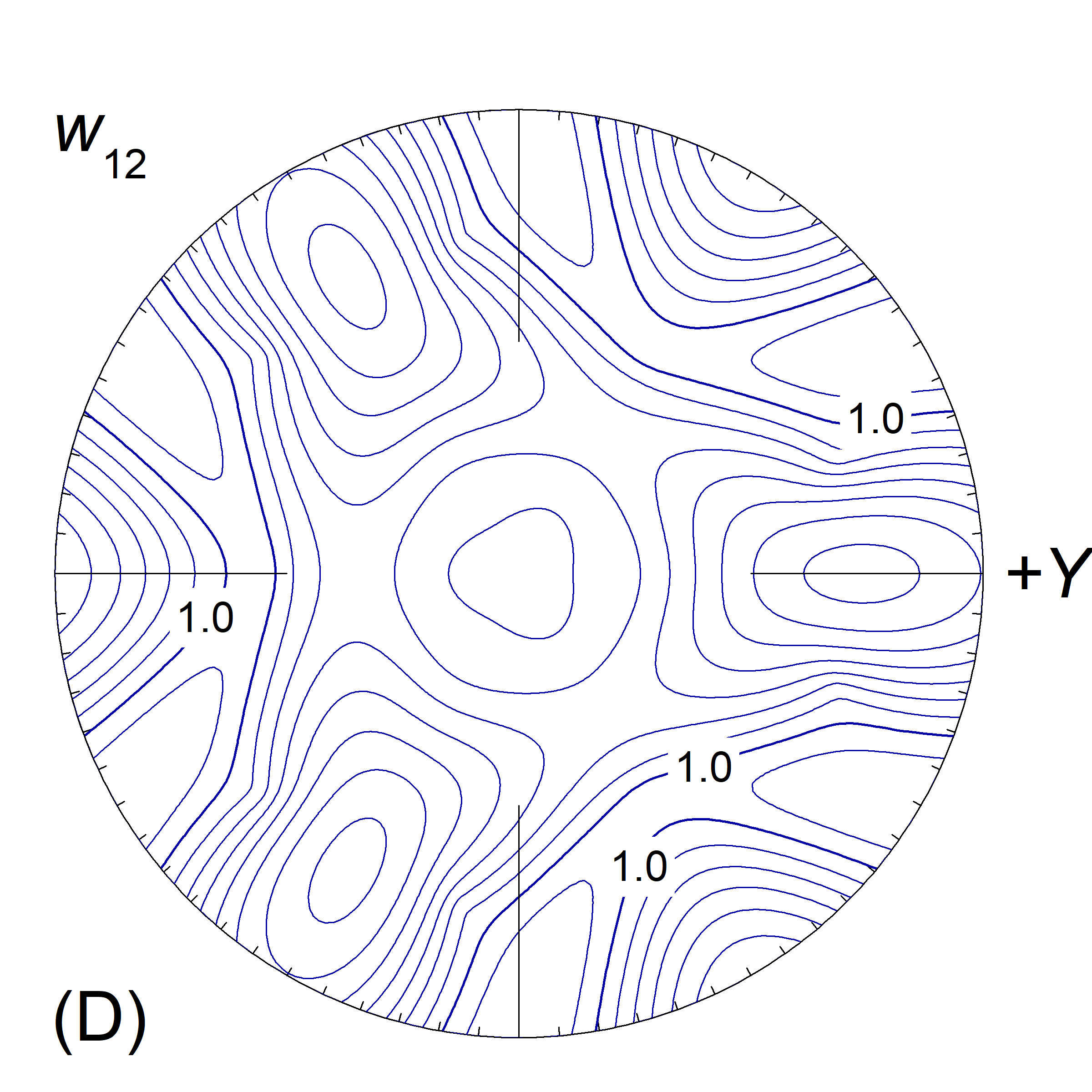}\includegraphics[width=0.25\textwidth]{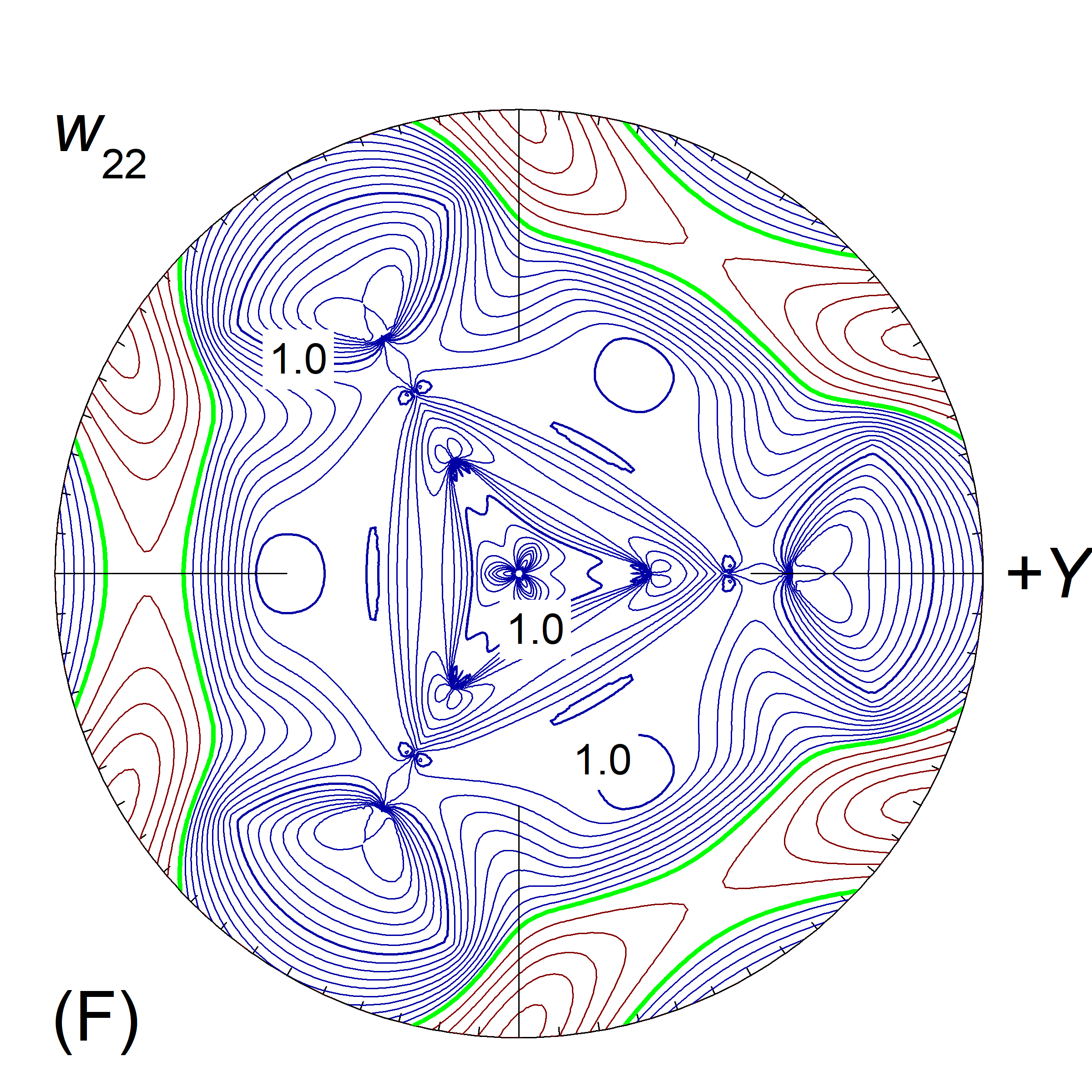}\includegraphics[width=0.25\textwidth]{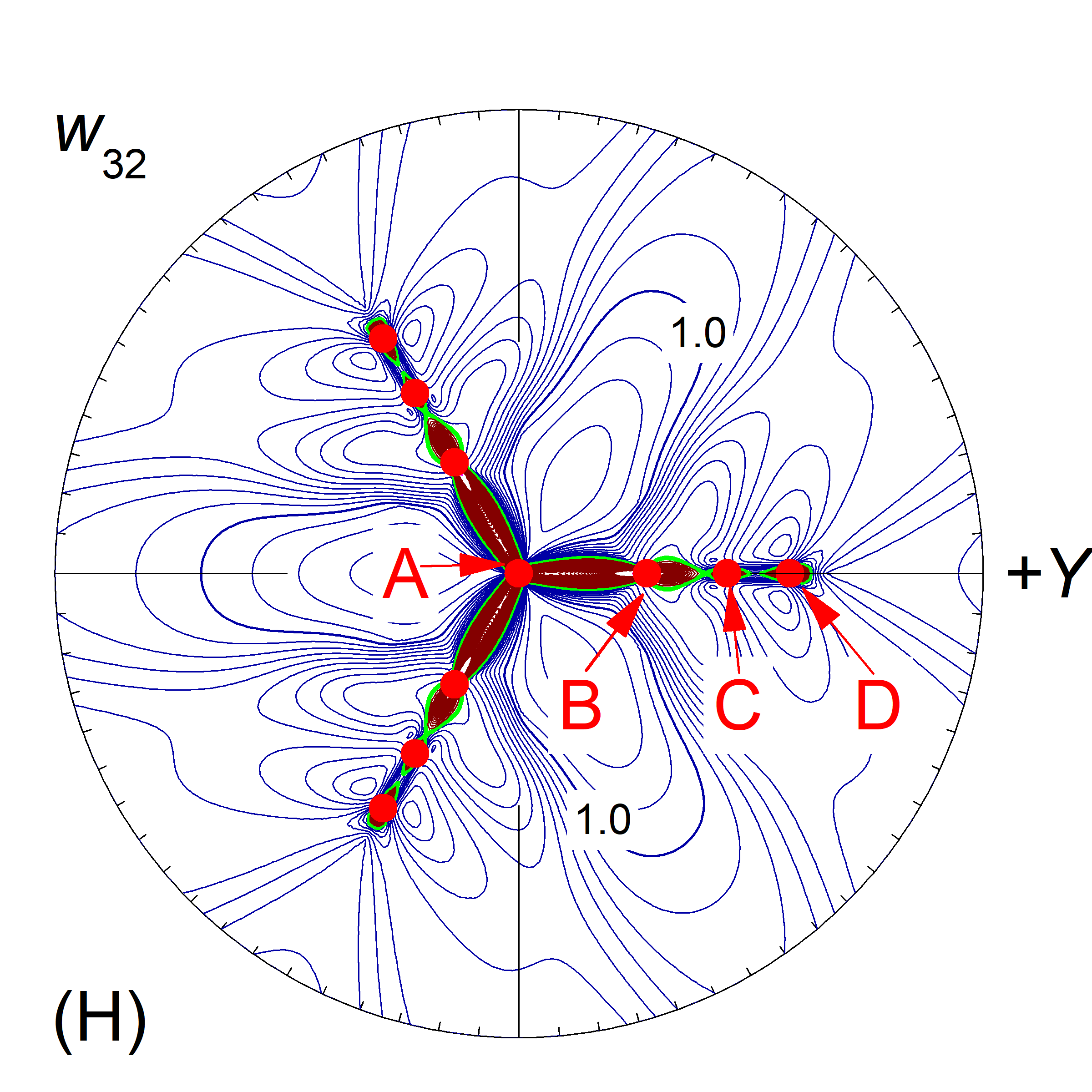}\\
  \caption{Acoustic diffraction tensor in lithium niobate crystal: (A) symmetry elements; (B) BAW axes and regions with high anisotropy $w_{21}w_{32}<-10$; (C-H) contour plots of BAW diffraction coefficients in stereographic projection (contour spacing is 0.1).}\label{fig-LNO}
\end{figure*}

\section{Analysis of selected crystals}\label{sec-res}

\subsection{Paratellurite}\label{sec-res-TeO2}
Paratellurite ($\alpha$-TeO$_2$) is a tetragonal crystal of 422 point group. 
The elastic constants of paratellurite and their temperature coefficients were measured by Silvestrova \emph{et al.}~\cite{SilvestrovaEtal87}. Paratellurite is the crystal with the strongest anisotropy of BAW properties~\cite{UchidaOhmachi69,LedbetterEtal04,BurovVoloshinovDmitrievPolikarpova11_eng}. The ratio of the maximum to the minimum BAW velocity is 7.2 and the energy flow angle reaches $74^\circ$. Piezoelectric effect in paratellurite does not affect acoustic anisotropy of the crystal, including the number and location of acoustic axes.

Stereographic projections of BAW diffraction coefficients in paratellurite are shown in Fig.~\ref{fig-TEO2}. Symmetry of the BAW slowness surface and its derivatives is 4/mmm. The high density of contour lines and the specific behavior of diffraction coefficients in the vicinity of acoustic axes reveal the locations of such axes in TeO$_2$. The positions of acoustic axes are marked with filled circles. Moreover, the areas characterized by fast variation of diffraction coefficients are also clearly seen. These areas connect acoustic axes and may include orientations with zero coefficients  $w_{\alpha2}$. Such orientations mean autocollimation of acoustic beam along one direction and can be very useful for application in acoustic or acousto-optic devices.

Acoustic axes in tetragonal crystals are restricted to symmetry planes~\cite{Khatkevich77_eng}. One of the axes is always associated with the 4-fold symmetry axis $Z$ and belongs to tangential type of degeneracy, label ``A'' in Fig.~\ref{fig-TEO2}. Therefore, all six coefficients $w_{\alpha\beta}$ in this direction are bounded. In $XY$ plane the slowness surface is decomposed to a circle and a quartic. The conical acoustic axes are the intersection points of the quartic outer sheet with the circle, label ``B'' (4 axes). Hereinafter, the axes obtained by multiplication with the symmetry elements of a crystal are labelled with the same letter and different subscripts.

One of specific features of acoustic anisotropy in TeO$_2$ is $c_{11} < c_{66}$ that was first mentioned by Uchida and Ohmachi~\cite{UchidaOhmachi69}. This leads to anomalous BAW degeneration, which involves the two fastest modes, in $XZ$ and $YZ$ planes.  A special type of BAW degeneration in tetragonal crystals takes place when $c_{11}=c_{66}$. In this particular case the quartic of slowness in $XY$ plane is a union of two ellipses with axes along directions $X+45^\circ$ and $X-45^\circ$.  Elastic constants $c_{11}=56.12$~GPa and $c_{66}=66.14$~GPa in paratellurite are rather close. Correspondent velocities of two fastest BAW modes are 3058 and 3320 m/s. A single acoustic axis which exists along the $X$ axis in the case $c_{11}=c_{66}$ is split into two symmetrical axes in $XZ$ plane, label ``C'' in Fig.~\ref{fig-TEO2} (4 axes). The diffraction coefficients $w_{11}$ and $|w_{22}|$ are high in the whole sector of $XZ$ plane between the anomalous axes. 

Thus, paratellurite has 9 acoustic axes, the theoretical maximum for tetragonal crystals, that agrees with theoretically predicted. {Due to strong acoustic anisotropy, EW lines in TeO$_2$  were found for all three bulk modes, including the fastest one~\cite{Naumenko83}.} Another peculiarity of BAW propagation in paratellurite is extremely low BAW velocity of 617 m/s for a pure shear BAW propagating along $X+45^\circ$ symmetry axis. The velocity quickly increases with the angle between $\mathbf n$ and the slowest direction leading to high but finite diffraction coefficients $w_{31}=50.9$ and $w_{32}=11.7$. Unlike acoustic axes, there is no discontinuity in $w_{3\beta}$ values.

\begin{figure*}[!t]
  \centering

  \includegraphics[width=0.25\textwidth]{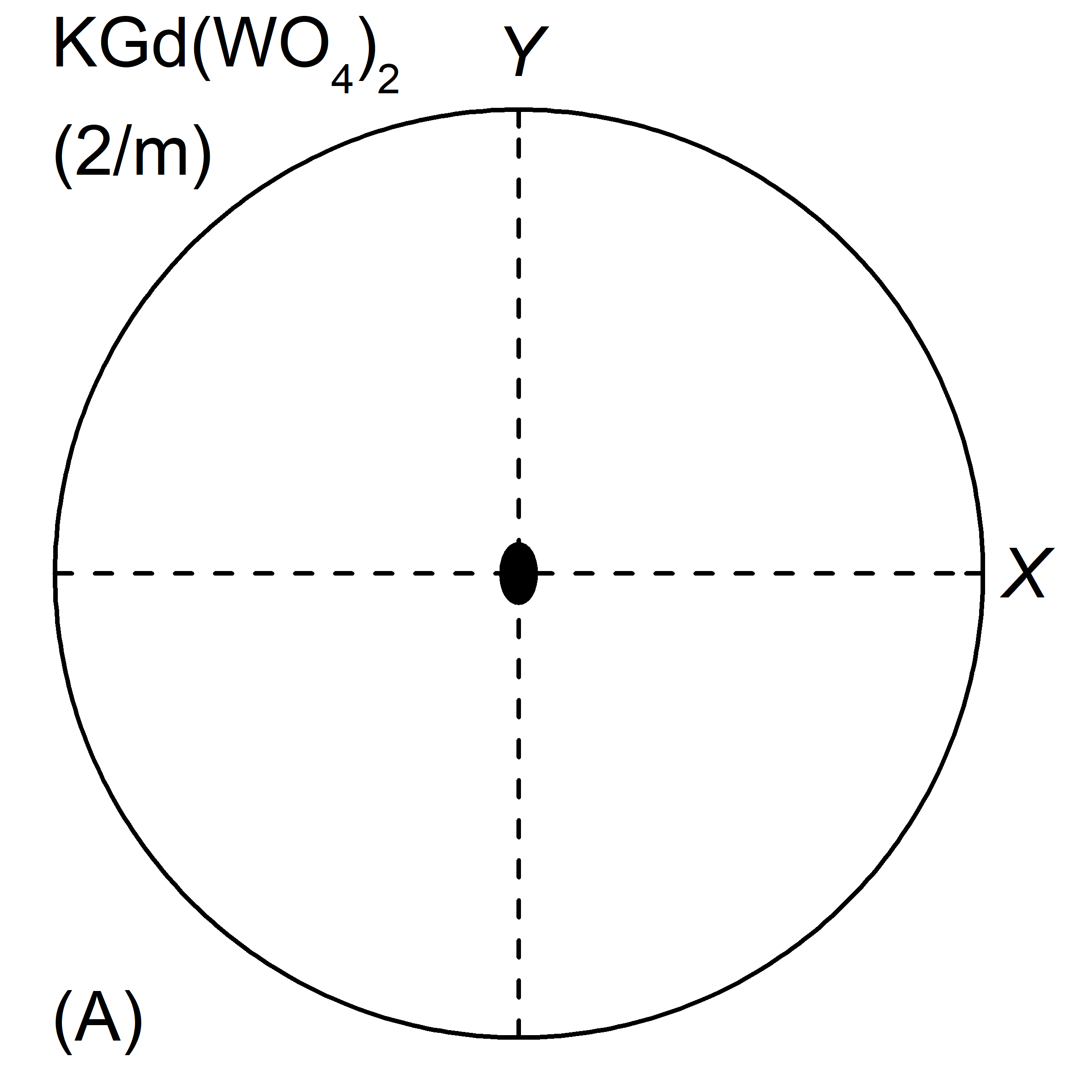}\includegraphics[width=0.25\textwidth]{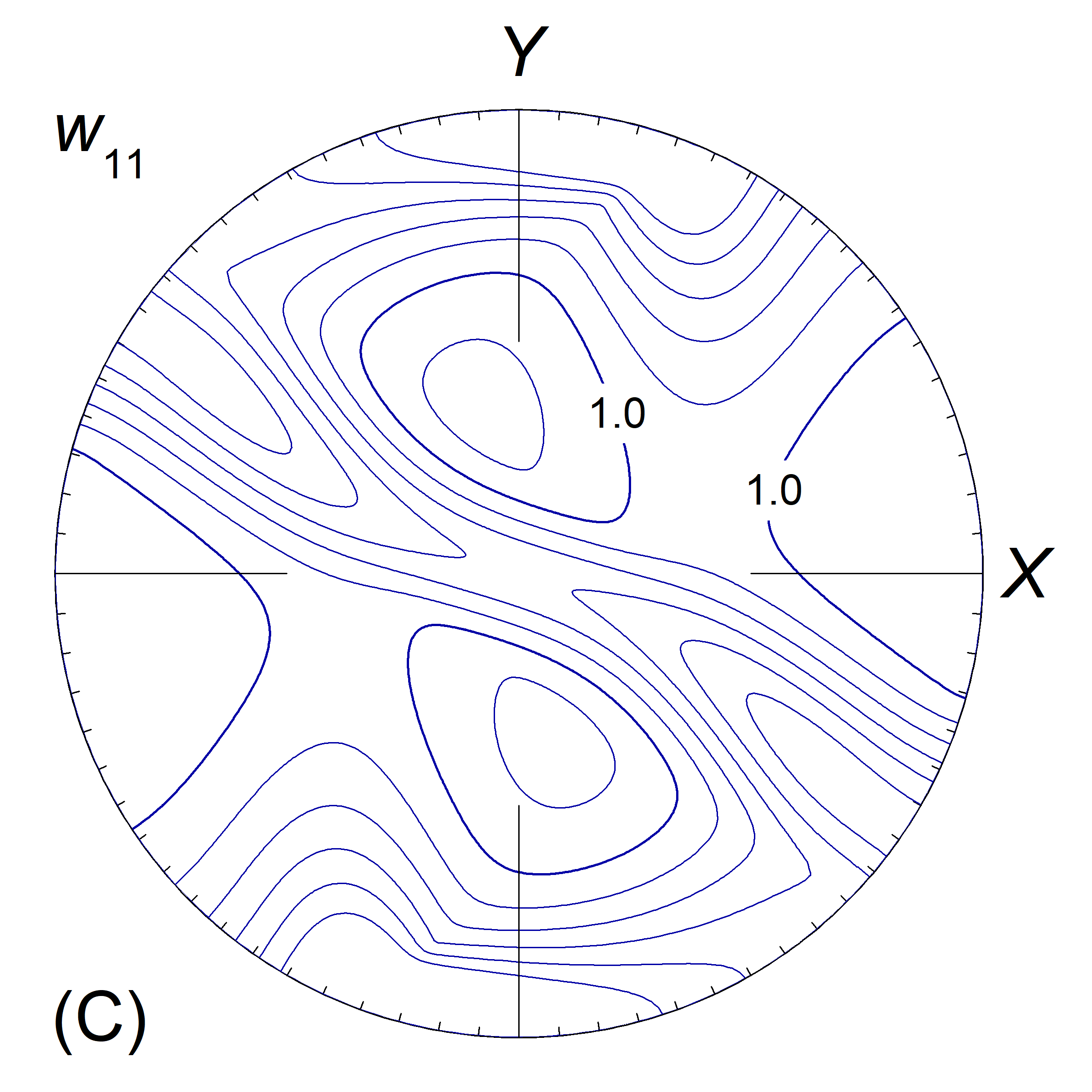}\includegraphics[width=0.25\textwidth]{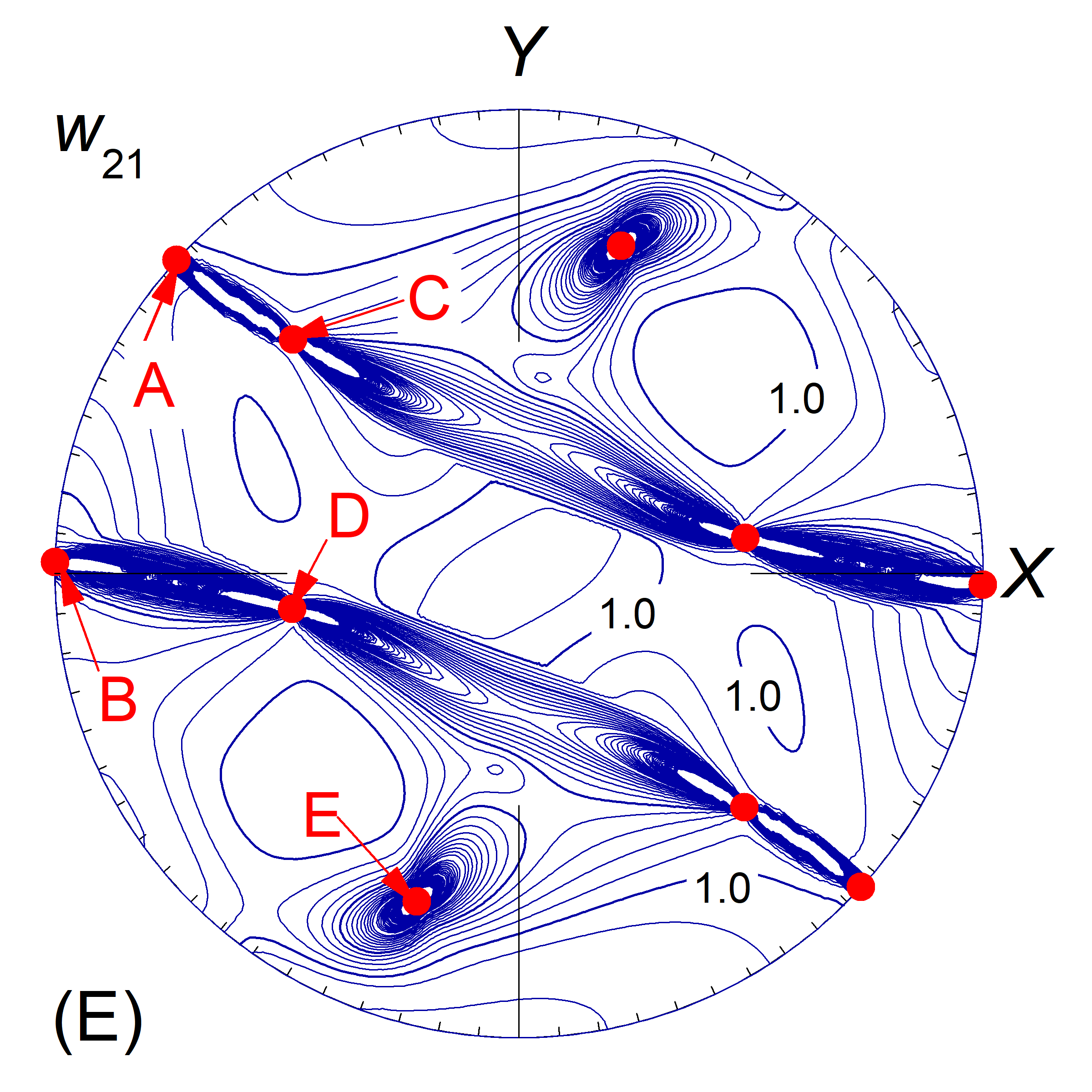}\includegraphics[width=0.25\textwidth]{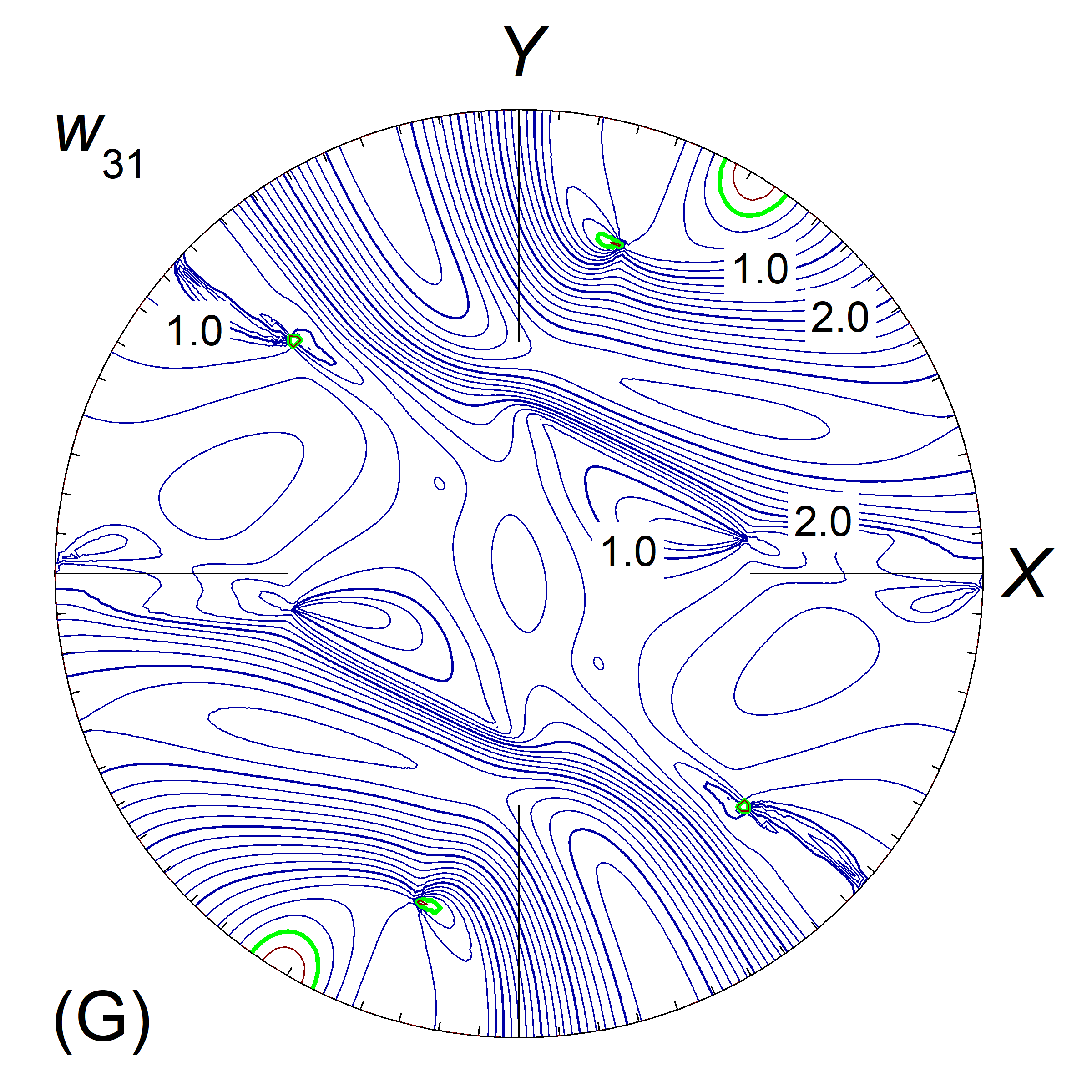}\\
  \includegraphics[width=0.25\textwidth]{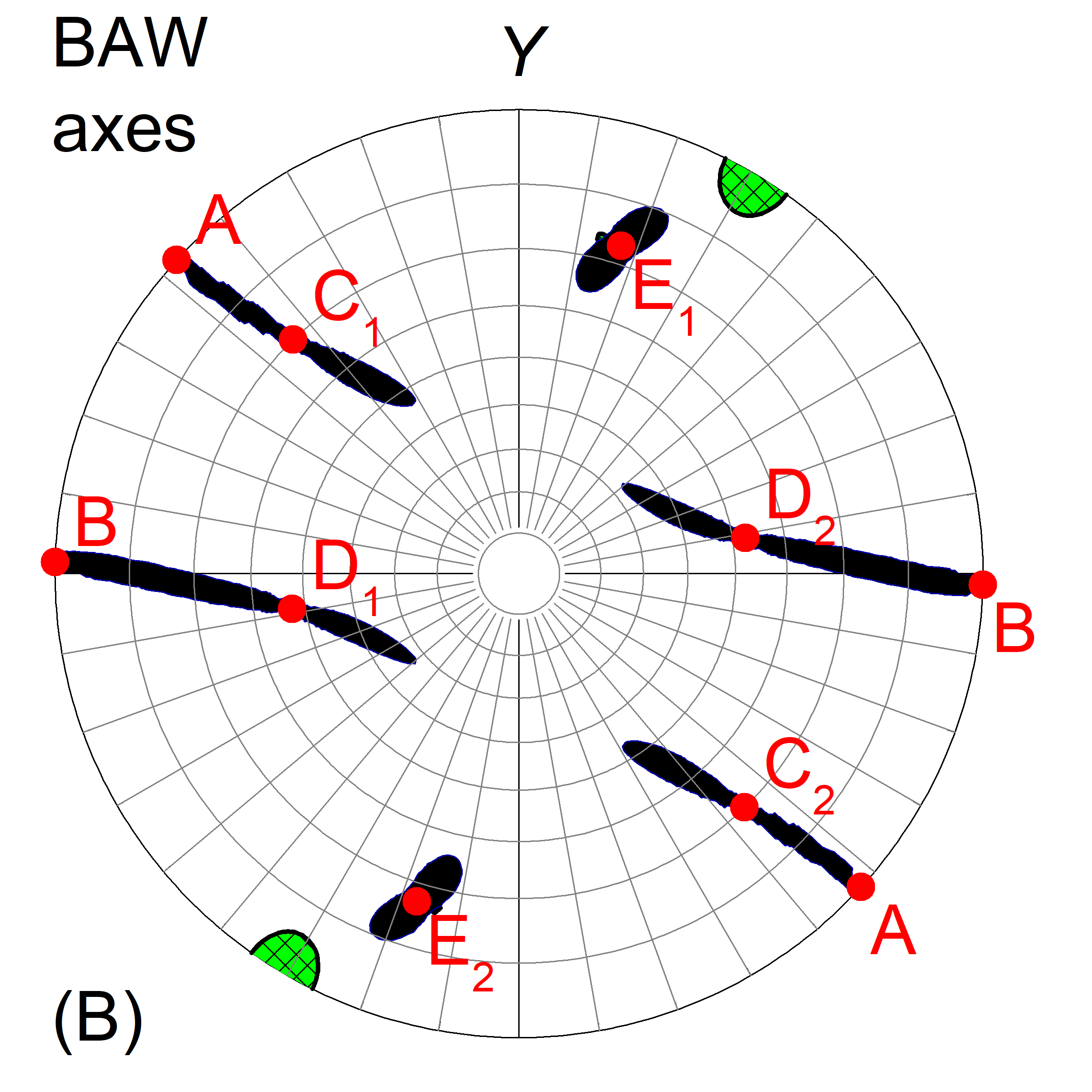}\includegraphics[width=0.25\textwidth]{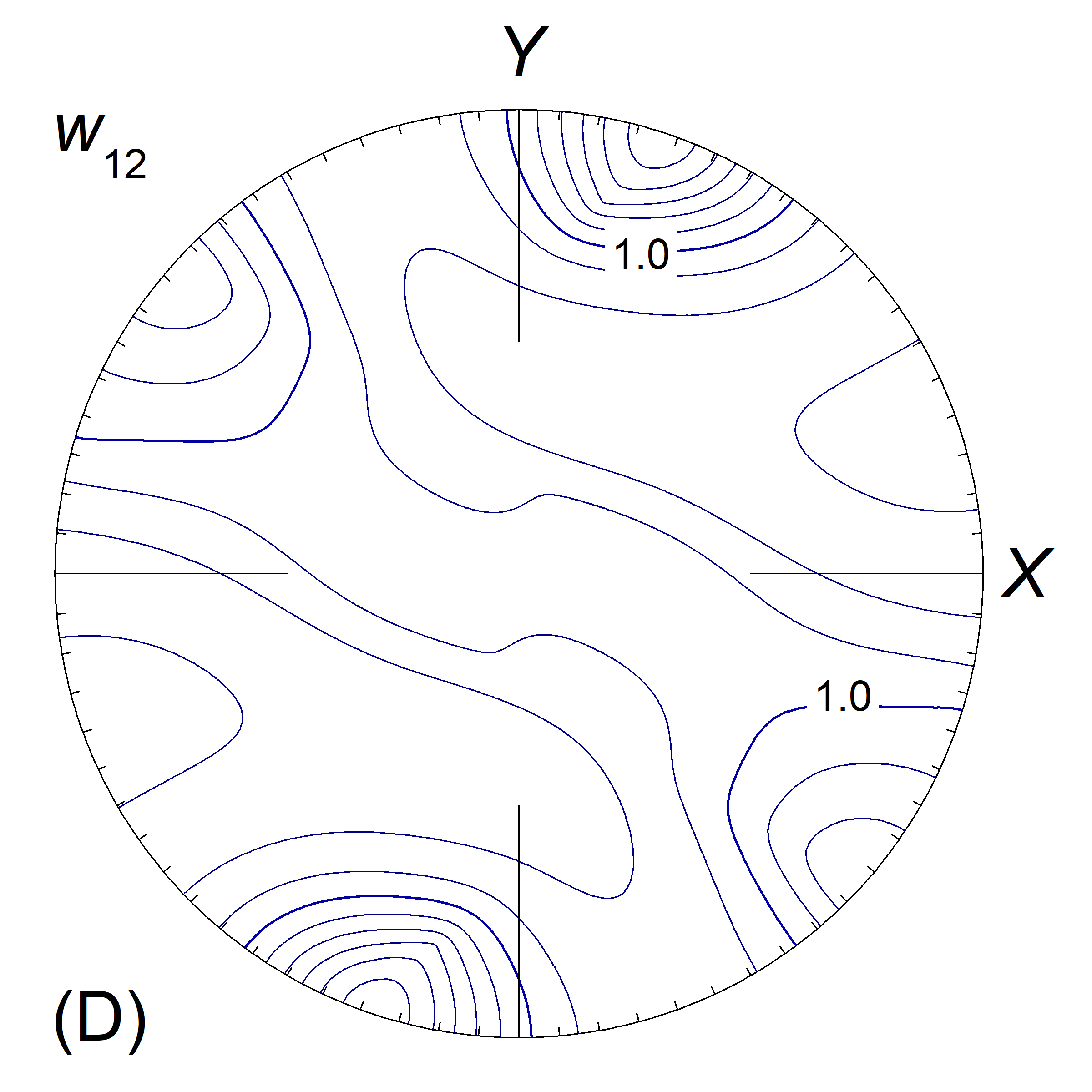}\includegraphics[width=0.25\textwidth]{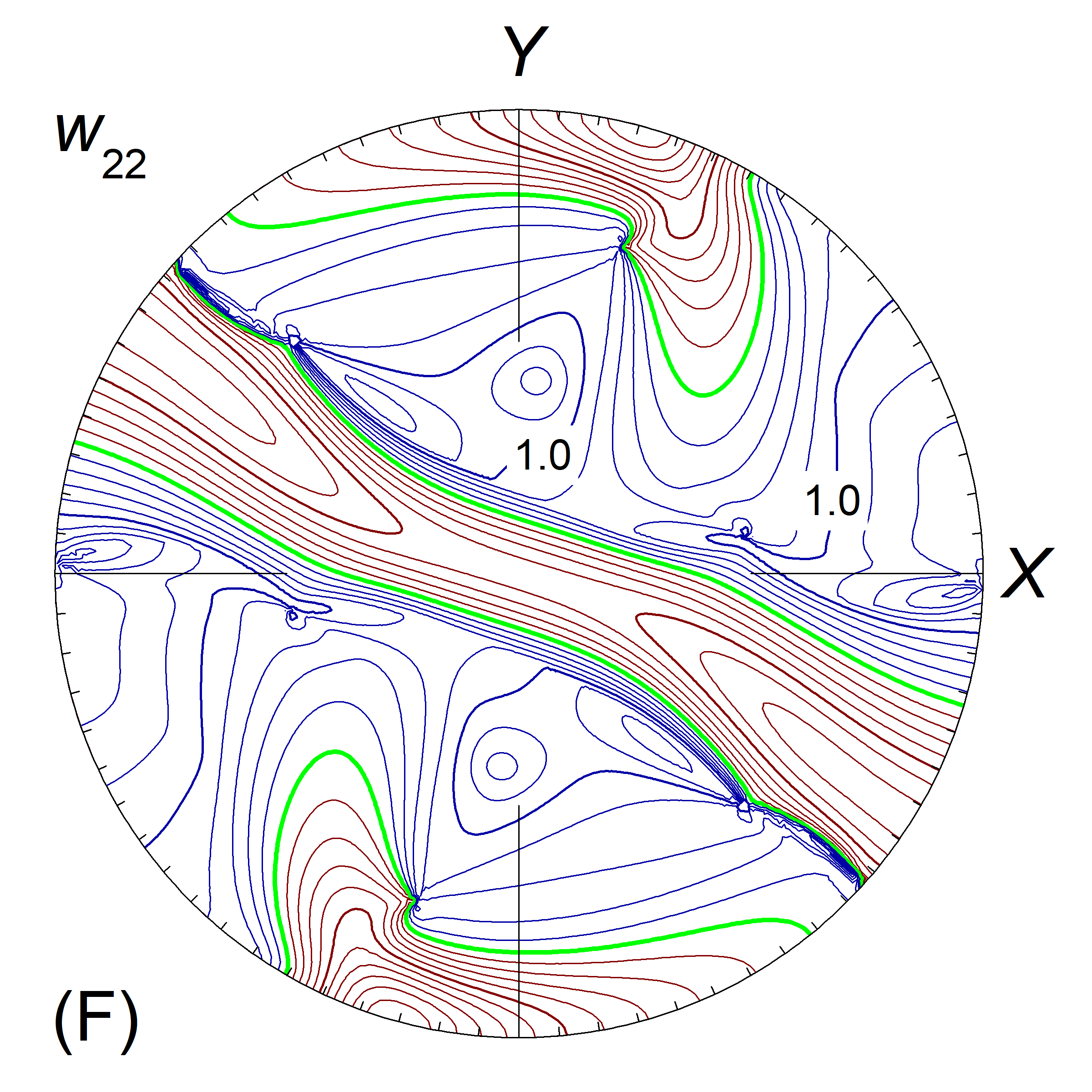}\includegraphics[width=0.25\textwidth]{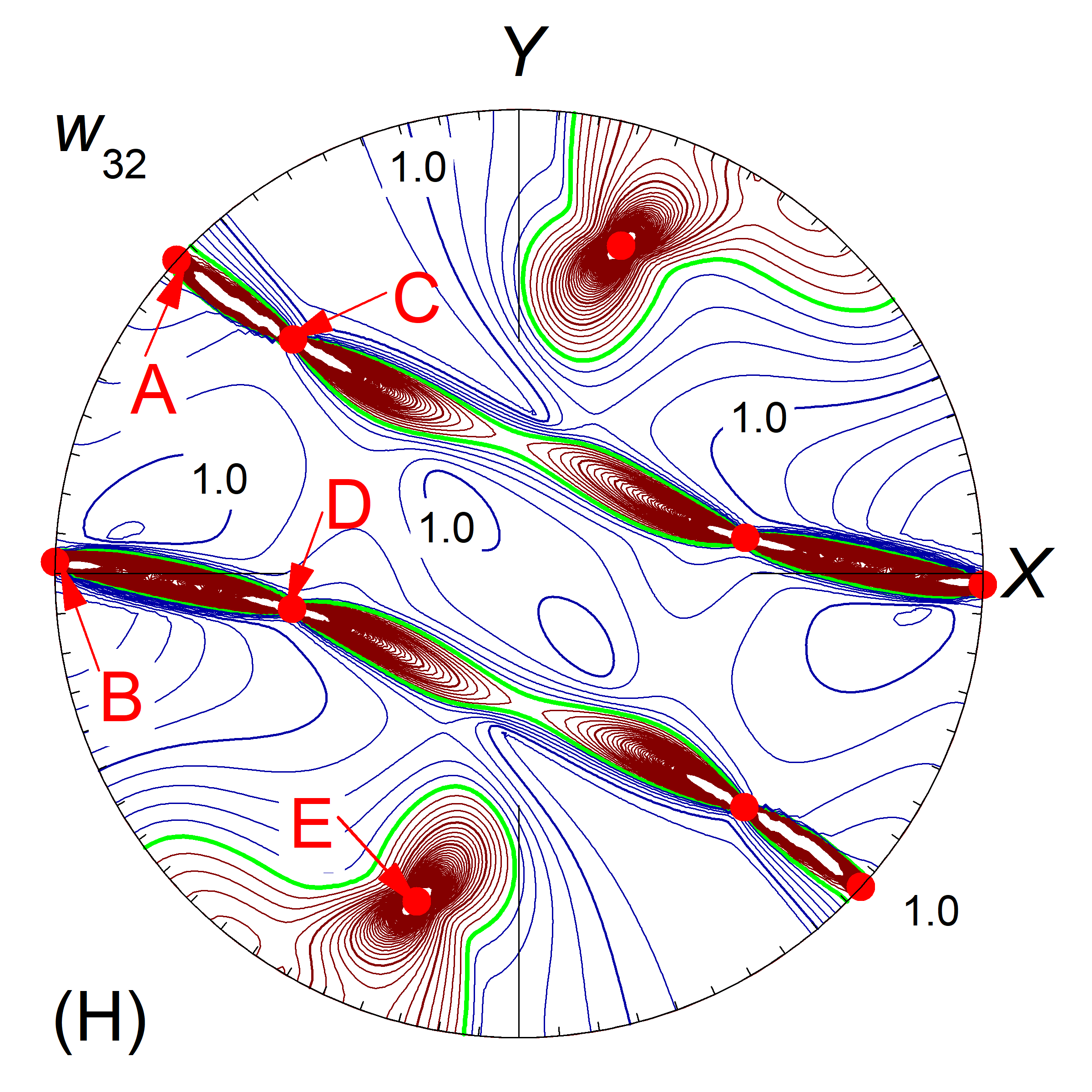}\\
  \caption{Acoustic diffraction tensor in KGW crystal: (A) symmetry elements; (B) BAW axes and regions with high anisotropy $w_{21}w_{32}<-10$; hatched regions indicate two-dimensional autocollimation of slow quasishear mode, $w_{31},w_{32}<0$; (C-H) contour plots of BAW diffraction coefficients in stereographic projection (contour spacing is 0.2).}\label{fig-KGW}
\end{figure*}

\subsection{Lithium niobate}

Lithium niobate (LNO, LiNbO$_3$) is a trigonal crystal of 3m point group. We use the material constants of LNO measured by Kushibiki \emph{et al.}~\cite{KushibikiTakanagaArakawaSannomiya99}. The results of simulations are shown in Fig.~\ref{fig-LNO}. Symmetry of the BAW slowness surface and its derivatives is $\overline{3}$m. Though acoustic anisotropy in LNO is not as strong as in paratellurite, zero contour lines (green lines in the figures) still can be found on the contour plots of diffraction coefficients $w_{22}$ and $w_{32}$.

The total number of acoustic axes is 10, with three axes in each crystal symmetry plane ($YZ$ and equivalent ones). Between the oblique axes there are no continuous regions of strong anisotropy. Moreover, there are autocollimation points, for the slow quasishear mode at the Euler angles $\theta = 44^\circ$ (between axes ``B'' and ``C'') and $\theta = 54^\circ$ (between axes ``C'' and ``D''). {There are no acoustic axes in this quadrant, and the energy flow angle reaches its maximum of $8.6^\circ$ that facilitates design of quasicollinear acousto-optic devices~\cite{NaumenkoMolchanovChizhikovYushkov15}}.

LNO is a piezoelectric crystal, which is most widely used due to very strong piezoelectric effect. It is a common material for high-frequency piezoelectric transducers for excitation of BAW waves in solids~\cite{GoutzoulisPape}. In the last years it found the main application in low-loss SAW filters and wireless SAW sensors. New application fields for this crystal appear due to increasing requirements to the new generation of SAW devices for communication systems and rapidly developing direct bonding technologies~\cite{HagelauerFattingerRuppel18,SahooOttavianoZheng18}. It allows combining LNO with quartz and other materials in a multilayered structure and assumes new optimization of crystal orientations based on good knowledge of its acoustic anisotropy~\cite{Naumenko19}. The multiple EW lines arising from acoustic axes give rise to the branch of low-attenuated leaky waves (see Sec.~\ref{sec-EWL}). Some of these leaky waves (e.g. propagating in $41^\circ$--$64^\circ$ $YX$-cuts) found wide application in low-loss SAW filters for mobile communication systems, due to combining of high velocity with high electromechanical coupling coefficient and low propagation losses.

\subsection{Potassium gadolinium tungstate (KGW)}

Potassium gadolinium tungstate (KGW, KGd(WO$_4$)$_2$) is a monoclinic crystal of 2/m point group. We choose the coordinate axes $XYZ$ in the following setting: $Z||\mathbf{b}$ is  2-fold symmetry axis, $Y||\mathbf{c}$. Elastic constants of KGW were measured by Mazur \emph{et al.}~\cite{MazurVelikovskiyMazur14}.

KGW is a laser host and nonlinear optical crystal. The family of potassium rare-earth tungstates has been recently rediscovered as a group of promising acousto-optic materials owing to their high laser damage threshold, good mechanical and thermophysical properties, established growth technology and processability~\cite{MazurVelikovskiyMazur14,MazurPozhar15_eng,SPIE19_10899}.

\begin{table}[!t]
  \centering
  \caption{Acoustic axes in KGW crystal}\label{tab1}
  \begin{tabular}{lcccc}
    \hline
    \hline
    Label & \multicolumn{2}{c}{Euler angles}  & Number  \\
     & $\theta$ (deg) & $\varphi$ (deg) & of axes \\
    \hline
    A & 90& 137.5 & 1  \\
    B & 90& 178.6 & 1\\
    C & 70.0 & 134.0 & 2 \\
    D & 52.6 & 8.9 & 2 \\
    E & 73.0& 72.7 & 2 \\
    \hline
    \hline
  \end{tabular}
\end{table}

Calculation results for KGW are shown in Fig.~\ref{fig-KGW}. There are 8 isolated acoustic axes. The axes positions are listed in Table~\ref{tab1}. Two axes labeled ``A'' and ``B'' are confined to $XY$ plane. The others are 6 off-plane axes, labels ``C'', ``D'', and ``E''. The plots for $w_{21}$ and $w_{32}$ show that there are continuous regions of high BAW divergence between axes ``A'' and ``C'' and between axes ``B'' and ``D''. Those regions are the ``ridges'' and ``valleys'' on the slowness surface, though the surface itself remains smooth and does not contain {other singularities between mentioned degeneracy points.}

Another peculiarity of BAW in KGW is existence of regions where $w_{31}<0$. This case is discussed in Sec.~\ref{sec-AC}.

\section{Discussion}

\subsection{Applications}
\subsubsection{Exceptional waves}\label{sec-EWL}

\begin{figure}
  \centering
  \includegraphics[width=0.5\columnwidth]{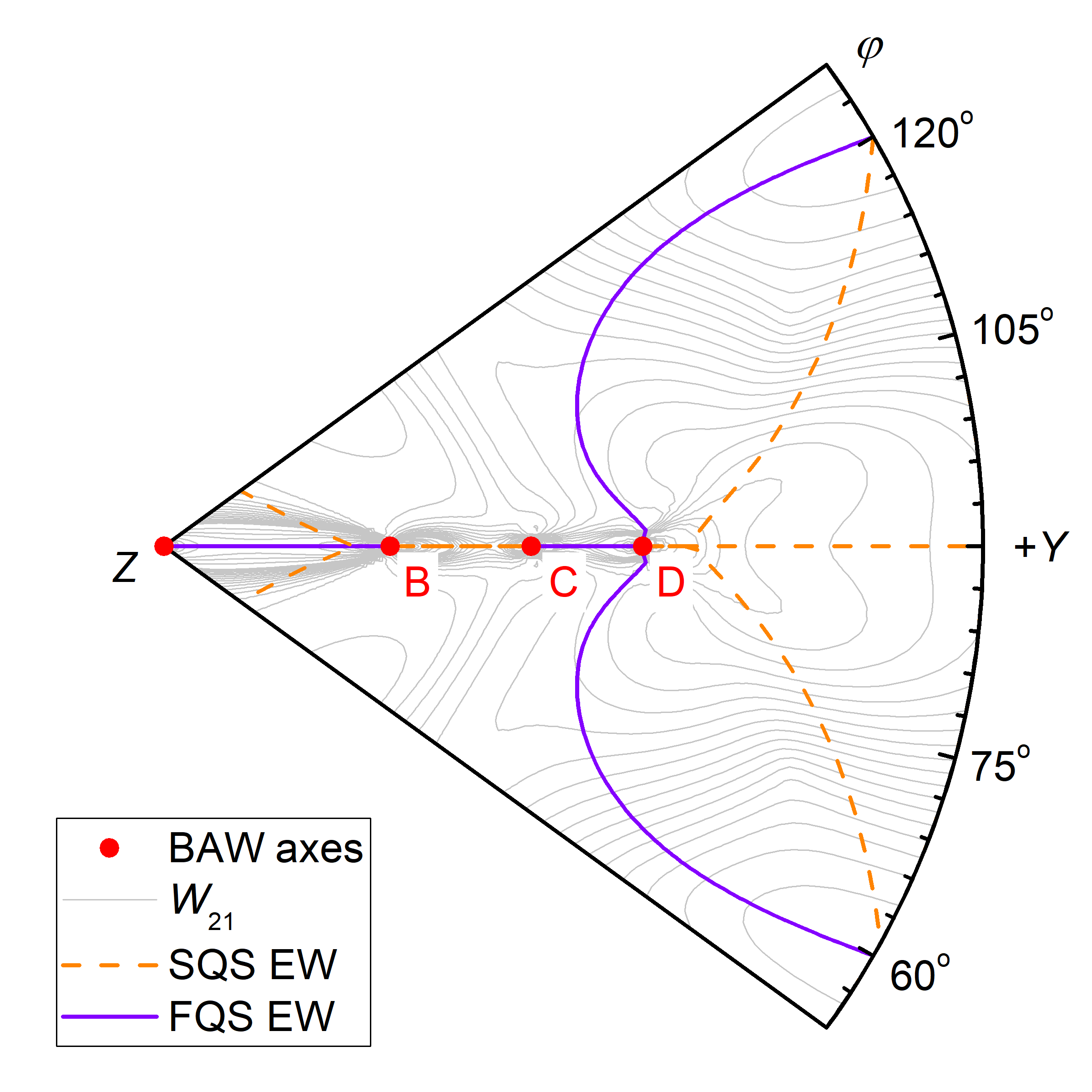}\\
  \caption{Exceptional wave lines in LNO crystal.​}\label{fig-SAW}
\end{figure}

Visualization of acoustic anisotropy helps solving different problems referred to design of BAW and SAW devices, for example, optimization of substrate orientations for high-frequency SAW devices. Fast variation of BAW velocities with propagation direction observed in the neighborhoods of acoustic axes is usually supplemented by fast rotation of polarization vectors. As a result, some BAWs can satisfy the stress-free mechanical boundary condition on selected crystal surfaces. Such ``exceptional'' BAWs can be slow shear, fast shear, or even longitudinal in some crystals. Propagation directions permitting existence of EWs form continuous lines on a stereographic projection of unit wave vectors (``EW lines''). In SAW devices, electrical boundary conditions and mass load of electrode structure modify the bulk wave into the quasi-bulk SAW or leaky SAW, but it usually stays low-attenuated and provides low propagation losses. Moreover, the branches of quasi-bulk low-attenuated leaky SAWs can arise from the fast shear and longitudinal EWs. Higher velocities of these waves, compared to common SAWs, facilitate fabrication of high-frequency SAW devices.

Figure~\ref{fig-SAW} shows a fragment of the stereographic projection of the BAW diffraction coefficient $w_{21}$ with added EW lines in LNO. Solid and dashed lines refer to the fast and slow quasishear and BAWs, respectively. Longitudinal EWs do not exist in this material. In addition to SH-polarized BAWs propagating in the symmetry plane YZ and switching between the fast and slow BAW modes in directions of four acoustic axes, three {off-plane} EW lines can be observed in the neighborhood of acoustic axes ``B'' and ``D''. Two lines arising in vicinity of the axis ``D'' cross $XY$ plane. The corresponding orientations are $38^{\circ}YX$ and $128^{\circ}YX$ cuts. $128^{\circ}YX$ cut supports propagation of the Rayleigh SAW and is widely used in SAW filters and sensors. $38^{\circ}YX$ cut gives rise to a branch of low-attenuated SH-type leaky SAWs. They propagate up to 1.25 faster than the Rayleigh SAW. The corresponding orientations are often applied in low-loss resonator SAW filters for communication systems.
\begin{figure*}[!t]
  \centering

  \includegraphics[width=0.25\textwidth]{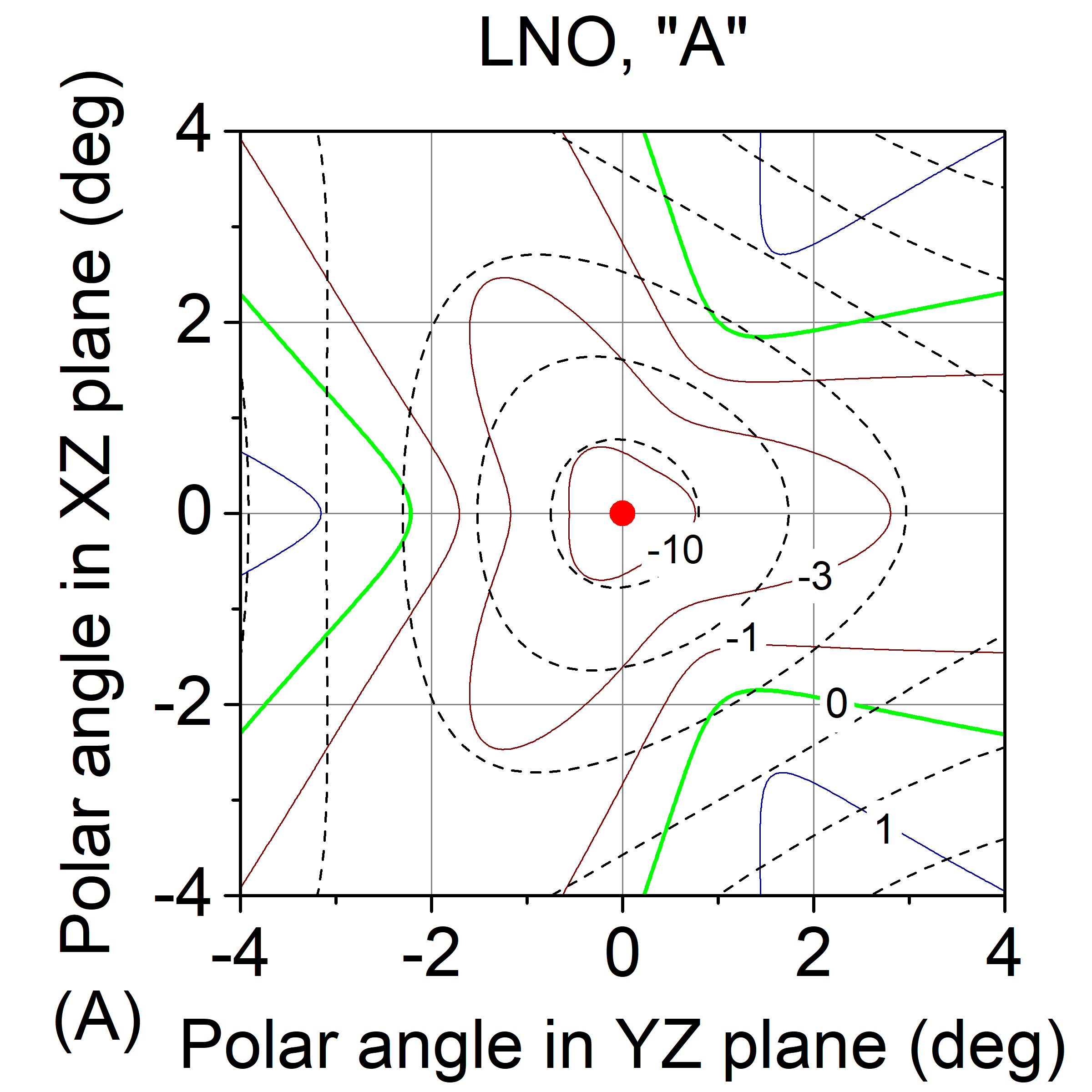}\includegraphics[width=0.25\textwidth]{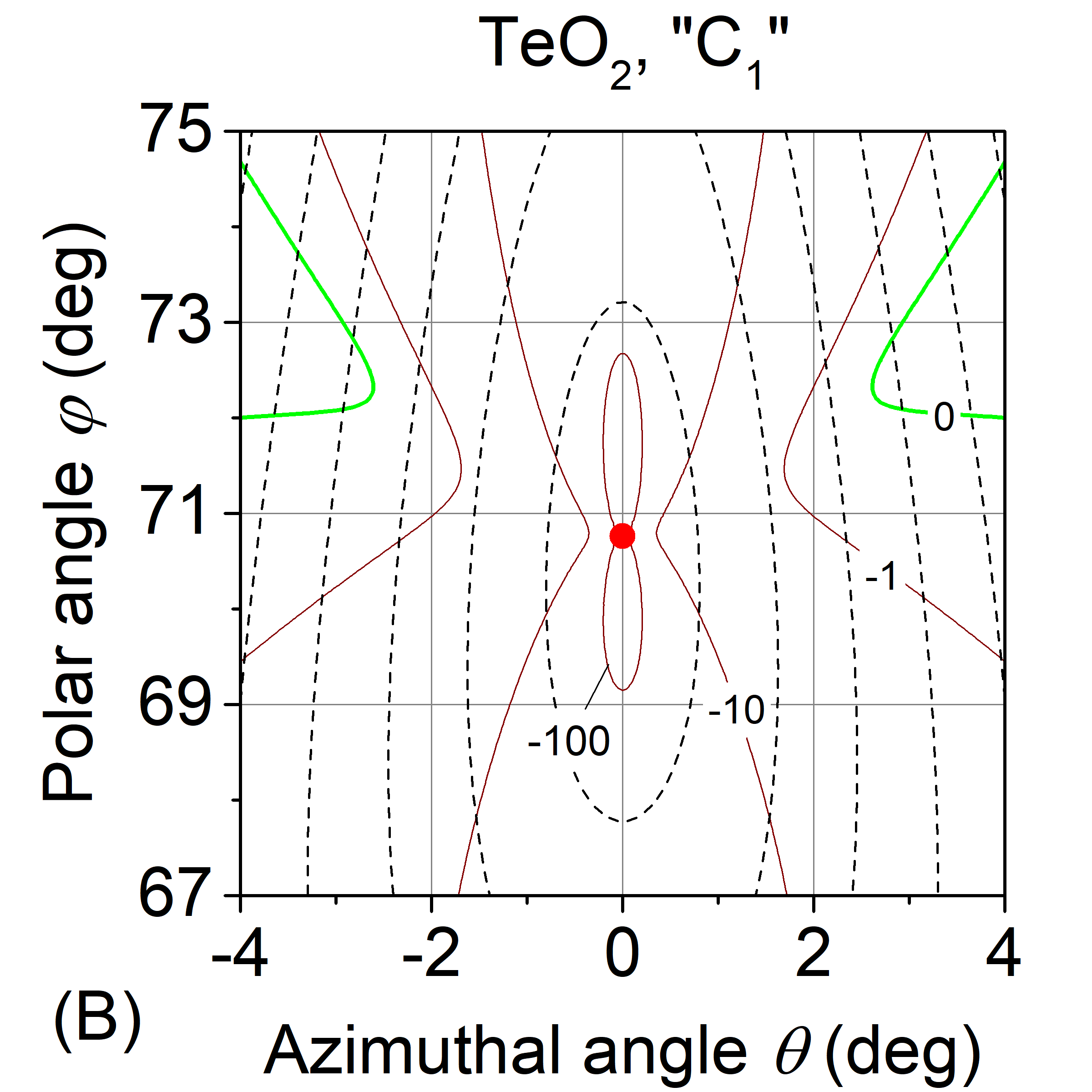}\includegraphics[width=0.25\textwidth]{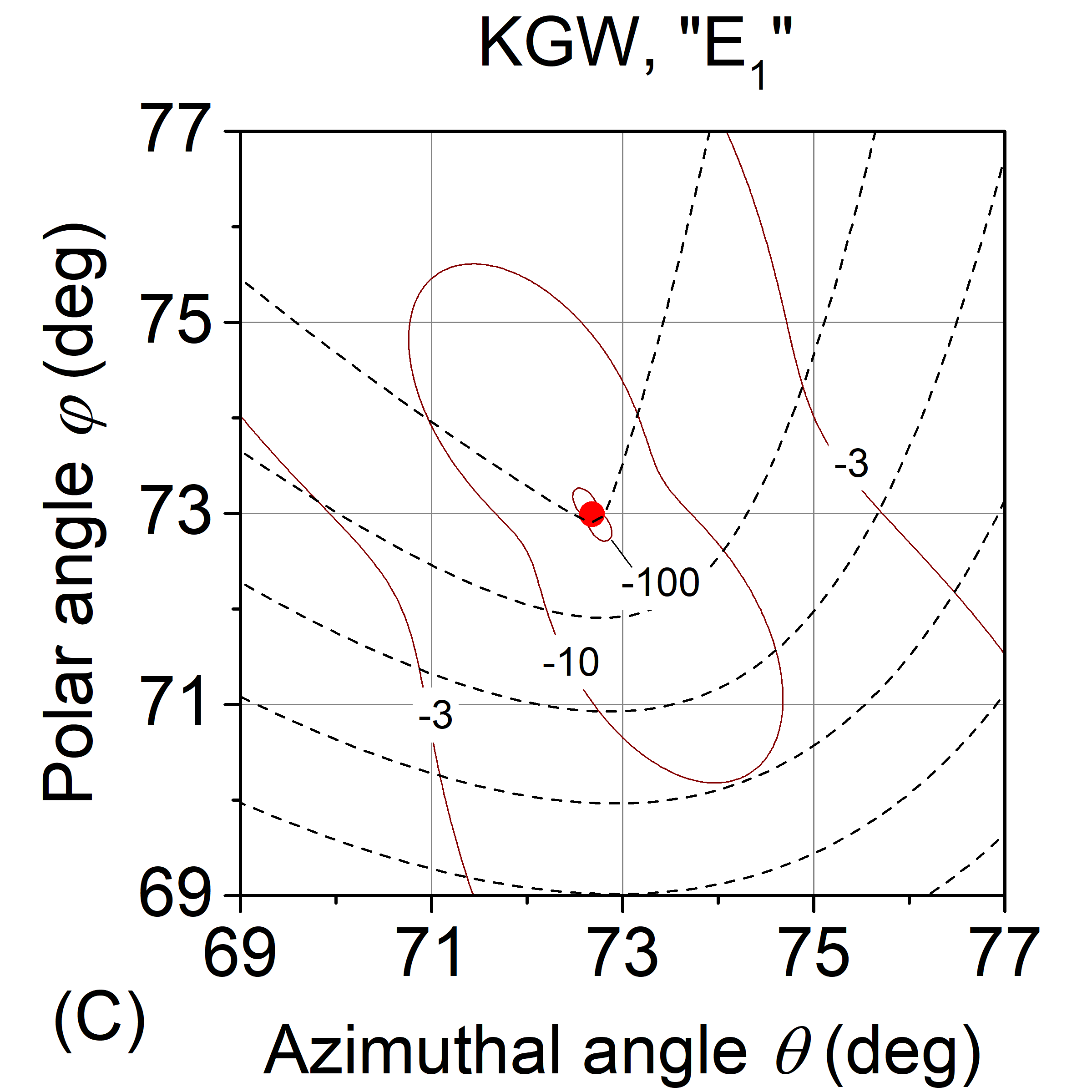}\includegraphics[width=0.25\textwidth]{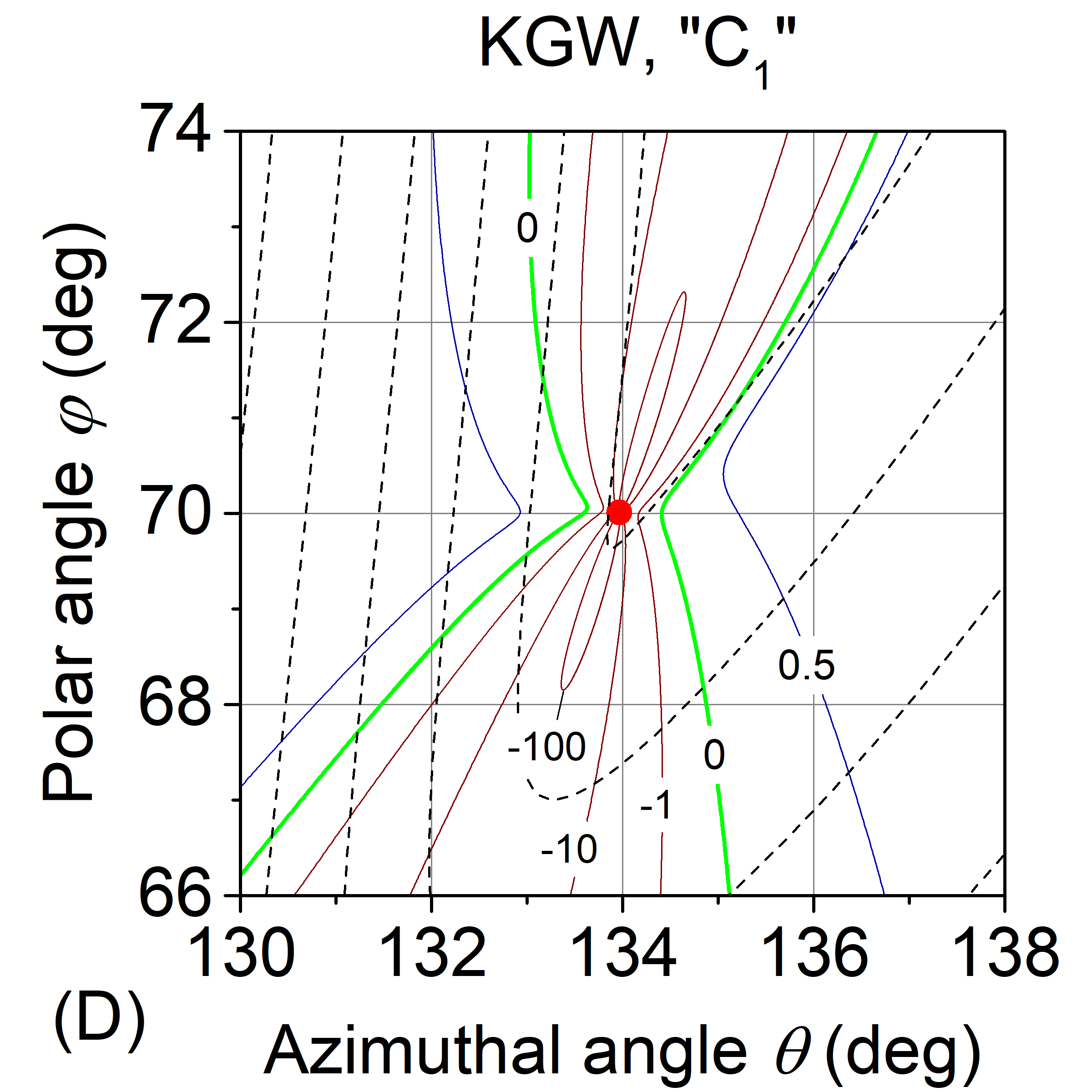}\\
  \caption{Isolines of BAW diffraction tensor eigenvalues (labelled solid lines) and slowness surface (dashed lines) in the neighborhood of conical acoustic axes: (A) $V_3$ and $w_{32}$ for 3-fold symmetry axis A axis in LNO; (B) $V_2$ and $w_{22}$ or anomalous C$_1$ axis in TeO$_2$; (C,D) $V_3$ and $w_{32}$ for off-plane E$_1$ and  C$_1$ axes in KGW.}\label{fig-topo}
\end{figure*}

Calculation and analysis of EW lines together with visualization of acoustic anisotropy can be an efficient tool in optimization of multilayered structures for SAW devices. Today such structures often replace regular SAW substrates in resonator SAW filters with improved quality factors. For example, a thin LNO plate bonded to quartz substrate provides a unique combination of high velocity, low attenuation, and high electromechanical coupling required for high performance SAW devices if orientations of both crystals are properly optimized based on analysis and visualization of their acoustic anisotropy. Non-attenuated quasi-longitudinal SAWs existing in these structures were found due to application of the method described here to investigation of acoustic anisotropy of quartz and calculations of EW lines in this crystal~\cite{Naumenko19}.

\subsubsection{Autocollimation directions}\label{sec-AC}

Among special directions of BAW propagation there are autocollimation directions where $w_{\alpha\beta}=0$. In such directions, self-diffraction of BAW beams is suppressed by anisotropy. Low beam divergency is used to reduce cross-talk in multichannel devices~\cite{KastelikEtal93,AubinSaprielMolchanov04}. Since any conical axis is associated with negative diffraction  tensor eigenvalues for the slower degenerate mode, autocollimation lines for this mode may exist close to the axis. {The behaviour of isolines near conical axes is  illustrated in Fig.~\ref{fig-topo} where the fragments of contour plots around conical axes are magnified.}

In a general case, autocollimation directions are not necessary related to acoustic axes. For example, Fig.~\ref{fig-KGW} demonstrates large BAW regions with $w_{\alpha\beta}<0$ in KGW crystal, including those for pure shear mode in $XY$ plane. For this mode $w_{22}<0$ between $X+40^\circ$ and $X+130^\circ$ directions, including the symmetry axis of the dielectric permittivity tensor $N_g$ ($X+101.5^\circ$) along which pure collinear Brillouin scattering takes place. Another configuration useful for practical applications is around the direction $X+60^\circ$ in $XY$ plane where both diffraction coefficients for slow quasishear wave polarized in $XY$ plane, $w_{31}$ and $w_{32}$, are negative. Those directions are specially marked in Fig.~\ref{fig-KGW}B. In this region, we predict almost spreading-free propagation of ultrasonic BAW beams because the diffraction coefficients are small, $w_{31}=-0.45$ and $w_{32}=-0.62$. An autocollimation point of the same kind for the slow quasishear wave in $XY$ plane exists in potassium yttrium tungstate (KY(WO$_4$)$_2$) crystals having similar properties to KGW~\cite{SPIE19_10899}.

\subsection{Diffraction tensor and other measures of anisotropy}

A close view on isolines of BAW diffraction tensor in a neighborhood of a conical acoustic axis (Fig.~\ref{fig-topo}) provides an insight of geometry of the slowness surface, which depends on crystal symmetry and position of the acoustic axis. {The isolines in an infinetesimal neighborhood of a conical axis have 2 orthogonal symmetry planes that complies with asymptotic representation of the slowness surface as an elliptical cone~\cite{Shuvalov98}. In the case of the 3-fold symmetry axis in LNO (Fig.~\ref{fig-topo}a) the isolines become circles.}

The method based on diffraction tensor allows finding of acoustic axes with higher accuracy than building of the slowness surface commonly used for visualization of acoustic anisotropy. If a conical acoustic axis does not coincide with a 3-fold symmetry axis (Fig.~\ref{fig-topo}A) but lies in a plane of elastic symmetry (Fig.~\ref{fig-topo}b) or takes general position in a low-symmetry crystal (Fig.~\ref{fig-topo}C and Fig.~\ref{fig-topo}D), the precise coordinates of such axis can be found with suggested numerical technique. Moreover, the areas of unusually strong acoustic anisotropy can be also clearly seen if they exist. For comparison, analytical approach helps to find positions of axes in symmetry planes but not outside them.

Eigenvalues of the diffraction tensor can be compared to anisotropy coefficients for BAW beams used by Kastelik \emph{et al.}~\cite{KastelikEtal93} and by Balakshy and Mantsevich~\cite{BalakshyMantsevich12AJ_eng}. The anisotropy coefficient is defined as the ratio of the beam ray angular spectrum width to the width of the plane wave angular spectrum. This phenomenological definition makes anisotropy coefficients dependent not only on the properties of the material, but also on configuration of the piezotransducer and acoustic frequency. Moreover, calculation of anisotropy coefficients is a highly resource-demanding numerical method based on computation and analysis of 3D structures of BAW beams for each direction in the crystal and each eigenwave. Calculations of the BAW diffraction tensor eigenvalues for paratellurite numerically coincide with anisotropy coefficients~\cite{KastelikEtal93,BalakshyMantsevich12AJ_eng}.

Finally, it should be noted that the eigenvalues of the diffraction tensor are a generalization of parabolic diffraction coefficient  (also called ``anisotropy parameter'') widely used in simulation and design of SAW devices. In SAW delay lines this parameter is responsible for diffraction losses. In resonator SAW filters the anisotropy parameter estimated in direction parallel to electrodes of an interdigital transducer (IDT) determines insertion losses and parasitic modes caused by coupling between SAW and transverse modes guided by IDT~\cite{NaumenkoAbbott04}. Similarly, diffraction losses of BAWs should be taken into account in design of other ultrasonic devices. One of important practical cases is a slow shear wave in paratellurite $X+45^\circ$ symmetry plane. This BAW mode is widely used in acousto-optic devices owing to its high coupling to optical waves. Recent experiments by Manstevich and Kostyleva revealed anomalous (i.e. non-exponential) attenuation of this wave~\cite{MantsevichKostyleva18}.

\section{Conclusions}

Contour plots of eigenvalues of BAW diffraction tensor provide fast and robust method for survey and visualization of acoustic anisotropy {of crystals, including a search for  singular directions (acoustic axes),} areas of fast or slow variation of BAW velocities, and directions where autocollimation of acoustic beams is expected. The number and position of BAW axes obtained using the diffraction tensor fully complies with theoretical predictions. Besides that, the proposed visualization method provides a comprehensive picture of BAW anisotropy in a crystal not only highlighting the degeneracy directions. {We demonstrated that analysis of crystal acoustic anisotropy is closely related to applications since} this method helps to locate crystallographic areas with characteristics required for a wide range of ultrasonic, acousto-electronic, and acousto-optic devices. This facilitates further detailed optimization of crystal orientation for these devices.


\section*{Acknowledgments}

The research was supported by the Ministry of Education and Science of the Russian Federation (Project 02.A03.21.0004 / Grant K2-2017-079) and the Russian Foundation for Basic Research (Project 17-07-00279).

\end{document}